\documentclass[twocolumn]{aastex62}

\usepackage{graphicx}
\usepackage{graphbox}
\usepackage{tikz}
\usepackage{float}
\usepackage{multirow}
\usepackage{color}
\usepackage[colorinlistoftodos]{todonotes}

\usepackage{float}
\usepackage{array, booktabs, makecell}
\usepackage{afterpage}

\accepted{to AJ}

\begin{document}
\large

    \title{ Alfnoor: A Retrieval Simulation of the Ariel Target List.} 
    \correspondingauthor{Q. Changeat}
	\email{quentin.changeat.18@ucl.ac.uk}
	\author[0000-0001-6516-4493]{Q. Changeat}
	\affil{Department of Physics and Astronomy \\
		University College London \\
		Gower Street,WC1E 6BT London, United Kingdom}
	\author[0000-0002-4205-5267]{A. Al-Refaie}
	\affil{Department of Physics and Astronomy \\
		University College London \\
		Gower Street,WC1E 6BT London, United Kingdom}
	\author[0000-0002-9007-9802]{L. V. Mugnai}
	\affil{Dipartimento di Fisica \\
		La Sapienza Universita di Roma \\
		Piazzale Aldo Moro 2, 00185 Roma, Italy}
	\author[0000-0002-5494-3237]{B. Edwards}
	\affil{Department of Physics and Astronomy \\
		University College London \\
		Gower Street,WC1E 6BT London, United Kingdom}
	\author[0000-0002-4205-5267]{I. P. Waldmann}
	\affil{Department of Physics and Astronomy \\
		University College London \\
		Gower Street,WC1E 6BT London, United Kingdom}
	\author[0000-0002-3242-8154]{E. Pascale}
	\affil{Dipartimento di Fisica \\
		La Sapienza Universita di Roma \\
		Piazzale Aldo Moro 2, 00185 Roma, Italy}
	\author[0000-0001-6058-6654]{G. Tinetti}
	\affil{Department of Physics and Astronomy \\
		University College London \\
		Gower Street,WC1E 6BT London, United Kingdom}



\begin{abstract} 

\large

In this work, we present \emph{Alfnoor}, a dedicated tool optimised for population studies of exoplanet atmospheres. Alfnoor combines the latest version of the retrieval algorithm TauREx\,3,  with the instrument noise simulator ArielRad and enables  the simultaneous retrieval analysis of a large sample of exo-atmospheres. We applied this tool to the Ariel list of planetary candidates and focus on hydrogen dominated, cloudy atmospheres observed in transit with the  Tier-2 mode (medium Ariel resolution). 

As a first experiment, we randomised the abundances -- ranging from 10$^{-7}$ to 10$^{-2}$ -- of the trace gases, which include H$_2$O, CH$_4$, CO, CO$_2$ and NH$_3$. This exercise allowed to estimate the detection limits for Ariel Tier-2 and Tier-3 modes when clouds are present. In a second experiment, we imposed  an arbitrary trend between a chemical species and the effective temperature of the planet. A last experiment was run requiring molecular abundances being dictated by equilibrium chemistry at a certain temperature. 

Our results  demonstrate the ability of Ariel Tier-2 and Tier-3 surveys to reveal trends between the chemistry and associated planetary parameters. 
Future work will focus on eclipse data, on atmospheres heavier than hydrogen and will be applied also to other observatories.
\end{abstract}

\section{Introduction} 

In the last decade, the field of extra-solar planets has very rapidly grown and matured. The NASA Kepler mission and other dedicated surveys from the ground have revolutionised our understanding of these extraterrestrial worlds. We are now aware of the ubiquity and vast diversity of planets outside our solar system, ranging from ultra-hot giant planets \citep{Gaudi_2017_kelt9, Delrez_2016_w121,Cameron_2010_w33} to more temperate Earths and Super-Earths \citep{Gillon_2016_trappist,Ment_LHS}. With TESS \citep{Ricker_2014_tess}, GAIA \citep{GAIA} Cheops \citep{Broeg_cheops_2013}, SPHERE \citep{SPHERE}, GPI \citep{Macintosh_GPI}, Espresso \citep{Pepe_espresso} currently operating and space missions like  PLATO \citep{rauer_plato} and WFIRST \citep{bennett_WFIRST} soon to come online, the statistics of planets in our galaxy will evolve even further in the next decade.

Current studies of exoplanetary atmospheres have been largely conducted using general observatories from space -- Hubble Space Telescope and Spitzer Space Telescope -- or from the ground -- e.g. VLT-Crires, NASA-IRTF, TNT-Giano, VLT-SPHERE, Gemini-GPI, Subaru- -- and thus results are often sparse and only available for a limited number  of the discovered planets. As a result, most atmospheric retrieval studies  have focused so far on the analysis of  individual planets \citep{Line_2016_hd209, Tsiaras_2016_55cnc, Kreidberg_2014_w43, Tsiaras_2019_k2-18} with only a few papers having attempted a consistent spectral analysis of multiple targets \citep{Tsiaras_pop_study_trans, Pinhas_ten_HJ_clouds, Barstow_2016_10planet, Sing_2015_popstudy}. In the next decade, a new generation of  observatories from space and the ground and dedicated missions \citep{Gardner_2006_JWST, Edwards_twinkle, Tinetti_ariel, ELT_2007, Skidmore_2015_TMT} will come online, offering a broader spectral coverage,  higher signal-to-noise ratio (SNR) and the ability to study a significantly larger number of targets. The ESA-Ariel mission alone has been designed to deliver transit, eclipse and phase-curve spectra for hundreds of planets, providing, for the first time, the chance to conduct a statistically significant survey of exoplanet atmospheres \citep{edwards_ariel}.

In most fields of astronomy (supernovae, brown dwarfs, black holes), revolutions in our understanding of the main processes often came from the study of the statistical behaviour using large  samples as opposed to individual studies. As the next generation of space telescopes come online, we will reach this important step for exo-atmospheres  and it is therefore critical to be aware of the challenges associated with large scale studies.

In this paper, we describe our integrated algorithm, Alfnoor, which combines the open source atmospheric retrieval code TauREx\,3 \citep{al-refaie_taurex3} and the Ariel noise simulator ArielRad \citep{mugnai_Arielrad} with the aim to facilitate the spectral analysis and interpretation of populations of exoplanetary atmospheres (\S \ref{sec:method}). 
Current Ariel's strategy is to observe planets in accordance to a four tier structure, where the aim of the second tier (Tier-2) of observations is to extract the key atmospheric constituents  \citep{edwards_ariel}.  In this paper we simulated Ariel Tier-2 and Tier-3 performances for a large sample  of planets provided in \cite{edwards_ariel}. For the selected targets, different, randomised atmospheric compositions were assumed and  an automated retrieval analysis for each planet was performed. We then compared and discussed the results of the posterior distributions, as provided by the retrievals, to the ground-truth to assess Ariel's ability to recover accurately and precisely the abundances of the key trace-gases and identify arbitrary injected chemical trends (\S \ref{sec:results}).
Finally we discuss these results in light of new facilities coming on line soon and next steps needed  to progress further in our understanding of population studies (\S \ref{sec:discussion}).

\section{Methodology and Software description} \label{sec:method}

\subsection{Description of the software}

To study large samples of exoplanetary spectra, we built a new tool: Alfnoor. Alfnoor combines the highly flexible next generation retrieval code TauREx\,3 with the ArielRad noise simulator to provide a unique framework dedicated to the study of exoplanetary populations with Ariel.

TauREx\,3 \citep{al-refaie_taurex3} is the new version of TauREx \citep{Waldmann_taurex2, Waldmann_taurex1}. This complete rewrite takes the form of a library and is designed to make customisation and external code integration easy. It uses the highly accurate line-lists from the ExoMol \citep{Tennyson_exomol}, HITRAN \citep{rothman} and HITEMP \citep{gordon} databases to build forward and retrieval models. A large number of options are available in terms of forward models (transmission, emission), chemical profiles (constant  as a function of pressure, two-layer, equilibrium chemistry), temperature profiles (isothermal, NPoints, \cite{Guillot_TP_model}) and cloud parameterisations (Grey, \cite{Lee_haze_model},\cite{bohren_2008_bhmie}).

ArielRad \citep{mugnai_Arielrad} estimates Ariel performances to observe a certain target when stellar, planetary and orbital parameters are specified. It also calculates the required number of observations to match the requirements for each of Ariel's tiers \citep{edwards_ariel}. In our study we focused on Tier-2 observations, which is the core of the  mission, and aims at characterising the key chemical species, thermal structure and the cloud properties of the selected atmospheres.  Ariel observations are expected to cover the wavelengths from 0.5$\mu m$ to 7.8$\mu m$. The telescope has 3 photometers: a Visible Photometer (VISPhot) and two Fine Guidance Sensors (FGS1 and FGS2) that are also used for the observations. The telescope also has two spectrometers: the Near Infrared Spectrometer (NIRSpec) and the Ariel Infrared Spectrometer (AIRS). The resolution of the spectrometers is adapted to the Tier levels. A description of the resolution achieved for each Tier can be found in \cite{Tinetti_ariel}, \cite{edwards_ariel} and \cite{mugnai_Arielrad}. It is summarised in Table \ref{tab:tiers}.

\begin{table}
$\begin{array}{|c|c|c|c|c}
\mbox{Instrument} & \mbox{$\lambda$ ($\mu$m)} & \mbox{R - Tier 1} & \mbox{R - Tier 2} & \mbox{R - Tier 3}\\
\hline
\mbox{VISPhot} & \mbox{0.5 - 0.6} & \mbox{$\O$} & \mbox{$\O$} & \mbox{$\O$} \\
\mbox{FGS1} & \mbox{0.6 - 0.8} & \mbox{$\O$} & \mbox{$\O$} & \mbox{$\O$} \\
\mbox{FGS2} & \mbox{0.8 - 1.1} & \mbox{$\O$} & \mbox{$\O$} & \mbox{$\O$} \\
\mbox{NIRSpec} & \mbox{1.1 - 1.95} & \mbox{1} & \mbox{10} & \mbox{20} \\
\mbox{AIRS-CH0} & \mbox{1.95 - 3.9} & \mbox{3} & \mbox{50} & \mbox{100} \\
\mbox{AIRS-CH1} & \mbox{3.9 - 7.8} & \mbox{1} & \mbox{10} & \mbox{20} \\

\end{array}$
\caption{ Wavelength coverage ($\lambda$) and resolutions (R) of each spectrometers (NIRSpec, AIRS-CH0 and AIRS-CH1) for the Ariel tiers. We also show the photometers (VISPhot, FGS1 and FGS2).}
\label{tab:tiers}
\end{table}  

The function \emph{Alfnoor-forward}  simulated high-resolution transit spectra with TauREx\,3 for all the targets. Next it called ArielRad to calculate the Ariel error bars, wavelength bins and the number of required observations to reach Tier-2  performances for all the targets.  The function \emph{alfnoor-inverse} took the Tier-2 spectra generated by alfnoor-forward and  performed  atmospheric retrievals using TauREx\,3 in fitting mode. 

Tier 1 observations  are studied in detail in \cite{Mugnai_2020_alfnoor1}.  Our sample of planets consists of the 146 planets observed in transit at Tier 2 from the Mission Reference Sample presented in \cite{edwards_ariel}. Of these planets, 14 of them qualify for observations in Tier 3. The simulated planets are built to represent the entire parameter space. In our sample 20 planets have radius smaller than 2 R$_E$, 29 are between 2 - 5 R$_E$ and 97 have radius $> 5$ R$_E$. For a more detailed description of the methodology used to build this target list, we refer the reader to \cite{edwards_ariel}. Future studies will concentrate on eclipse observations and / or secondary atmospheres.

\subsection{Approach and initial setups}
In all the models,  the atmosphere is composed of H$_2$ and He with a ratio  He/H$_2$ = 0.17.  For the trace-gases, the list and sources of the opacities used in this paper are presented in Table \ref{tab:fig_references}. Collision Induced Absorption for H$_2$-H$_2$ and H$_2$-He and Rayleigh scattering are included. For the retrievals, unless specified otherwise, we used the same assumptions:  mixing ratios constant with pressure, temperature constant with pressure, grey opaque clouds. While temperature variations with altitude are crucial for eclipse observations, in the case of transmission spectra, most studies assume isothermal temperature profiles. This is justified by the narrow wavelength coverage and signal-to-noise in available observations (with the Hubble Space Telescope) which is not allowing to probe large pressure regions in the planet atmosphere. The temperature variations in transmission act as a second order parameter and the spectrum is most sensitive to the mean temperature value, which directly appear in the scale height. However, \cite{barstow2013potential}, \cite{Rocchetto_biais_JWST} and \cite{Changeat_2019_2layer} highlighted the impact of temperature variations for high signal-to-noise and broad wavelength coverage cases, indicating that JWST and Ariel would be able to retrieve more complex temperature structures from transit spectra. As this study focuses on the capabilities of Ariel to recover chemical species, we do not investigate further the impact of non-isothermal temperature structures. We however note that this assumption could introduce biases to our results. Parameters that are traditionally determined using external methods are fixed to the true values: e.g. stellar radius, planetary mass and He/H$_2$ ratio. The list of free parameters along with the priors used are described in Table \ref{tab:fit_parameters}
\begin{table}
\centering

$\begin{array}{|c|c|}
\mbox{Opacity} & \mbox{References}  \\
\hline
\mbox{H}_2\mbox{-H}_2 & \mbox{\cite{abel_h2-h2}, \cite{fletcher_h2-h2}} \\
\mbox{H}_2\mbox{-He} & \mbox{\cite{abel_h2-he}} \\
\mbox{H}_2\mbox{O} & \mbox{\cite{barton_h2o}, \cite{polyansky_h2o}} \\
\mbox{CH}_4 & \mbox{\cite{hill_xsec}, \cite{exomol_ch4}} \\
\mbox{CO} & \mbox{\cite{li_co_2015}} \\
\mbox{CO}_2 & \mbox{ \cite{rothman_hitremp_2010}} \\
\mbox{NH}_3 & \thead{\mbox{ \cite{Yurchenko_nh3_2011},} \\ \mbox{\cite{tennyson2012exomol}} }\\

\end{array}$
\caption{List of opacities used in this work}
\label{tab:fig_references}
\end{table}

\begin{table}
\centering

$\begin{array}{|c|c|c}
\mbox{Parameters} & \mbox{Priors} & \mbox{Scale}  \\
\hline
\mbox{radius (R$_J$)} & \mbox{$\pm 50 \%$} & \mbox{linear} \\
\mbox{cloud pressure (bar)} & \mbox{$10$ - $10^{-7}$} & \mbox{log} \\
\mbox{T (K)} & \mbox{$\pm 50 \%$} & \mbox{linear} \\
\mbox{H$_2$O (VMR)} & \mbox{$10^{-12}$ - $10^{-1}$} & \mbox{log} \\
\mbox{CH$_4$ (VMR)} & \mbox{$10^{-12}$ - $10^{-1}$} & \mbox{log} \\
\mbox{CO (VMR)} & \mbox{$10^{-12}$ - $10^{-1}$} & \mbox{log} \\
\mbox{CO$_2$ (VMR)} & \mbox{$10^{-12}$ - $10^{-1}$} & \mbox{log} \\
\mbox{NH$_3$ (VMR)} & \mbox{$10^{-12}$ - $10^{-1}$} & \mbox{log} \\

\end{array}$
\caption{List of the fit parameters and their priors for the retrievals ({\it Alfnoor-inverse}). We take a conservative approach and select larger bounds than the ones used to randomly generate the planets in forward mode. The chemical abundances are expressed in Vertical Mixing Ratios (VMR).}
\label{tab:fit_parameters}
\end{table}

In this study, we aim to explore two particular aspects of the Ariel mission:
\begin{enumerate}
\item the ability of Ariel to detect molecular species and the detection limits for these molecules in the context of cloudy primary atmospheres observed in transit. This task can be easily achieved by performing retrievals on an unbiased  dataset of planets where the atmospheric composition is randomised and by assessing the cases that have been successfully recovered. 

\item  the ability of Ariel to reveal chemical trends in exoplanet populations. To assess this possibility,  a biased sample can be used as input where an artificial trend is introduced. 
\end{enumerate}
We describe below the actual implementation of this plan. 
\begin{enumerate}
\item \underline{Unbiased sample}. We  built the forward model by using the stellar and planetary basic  parameters from \cite{edwards_ariel} for the Ariel Target list.  We randomised the chemistry, temperature and cloud parameters so that a unique set of these parameters is adopted for each planet of each sample. For the chemistry, we considered  constant profiles with pressure for the mixing ratios of H$_2$O, CH$_4$, CO, CO$_2$ and NH$_3$ and chose a random abundance in logarithmic scale from $10^{-7}$ to $10^{-2}$. For clouds, we generated  grey opaque clouds with random top pressures varying in log-scale from 10 bar (equivalent to no clouds) to $10^{-3}$ bar. Finally, the atmospheric temperatures were also randomly generated and allowed to assume values between $0.7\times T_{eff}$ and $1.05\times T_{eff}$, where $T_{eff}$ is the effective temperature in the Ariel target list of \cite{edwards_ariel}. The temperature was consciously selected biased towards lower values to account for differences between effective temperature and the terminator temperature \citep{Caldas2019_3dsim}. 
 We repeated the generation of the observed spectra twice to build ‘un-scattered’ and ‘scattered’ datasets. In the un-scattered set, we conserve the theoretical simulated spectra as-is. As scatter generally arises from the random realisation of observations, we apply a Gaussian scatter to a second dataset using the true value as mean and the simulated noise as variance. This scattered dataset better describes what would be obtained in an actual observation by the telescope but cannot be used to characterise retrieval biases as ‘unfortunate’ runs could lead to large discrepancies between true and retrieved values \citep{Feng_retrieval_earthanalog, Changeat_2019_2layer}. Unscattered spectra are more suitable for the study of retrieval biases and intrinsic correlations between the atmospheric parameters \citep{Feng_retrieval_earthanalog}. On the opposite, scattered spectra can inform us on the stability and the redundancy in the information content of Ariel spectra. Previous studies have used both types to simulate observations by future telescopes \citep{barstow2013potential, Tinetti_2015_echo, Feng_2016_nonuniform_tp, Rocchetto_biais_JWST, Molliere_2017_jwstsim, Batalha_2017, Tinetti_ariel, Blumenthal_2018_jwst, Feng_retrieval_earthanalog, Edwards_twinkle, Changeat_2019_2layer, Lustig_Yaeger_2019_trappist, changeat_2020_mass}. \cite{Feng_retrieval_earthanalog} results predicted that the retrieved uncertainties should be similar in both scattered and un-scattered runs but that the retrieved mean could be different. Here, we use our two datasets to investigate these predictions keeping in mind that if Ariel spectra contain enough information content redundancy, we should not see large differences in the retrieved mean values. 

\item \underline{Biased samples}. We imposed first a linear relationship between the logarithmic abundance of water and the temperature. We enacted this correlation water-temperature by requiring a mixing ratio of $10^{-4}$ for an effective temperature of 1000K and $10^{-3}$ for an effective temperature of 2000K.

We then tried a more realistic example where the atmospheres were assumed to be in chemical equilibrium and simulated accordingly the chemical abundances and profiles  \citep{Agundez_2012_eqchem}.  We used the same solar C/O ratio and metallicity for all the planets in the sample. To recover the input profiles, we used in the retrievals both free, constant with altitude chemical profiles and profiles which are forced to follow chemical equilibrium prescriptions. 
We did not test  the entire sample with the two-layer chemistry retrieval scheme as presented in \cite{Changeat_2019_2layer}, but we have run an example to show the expected improvements of this scheme over the  pressure constant chemical profiles.
\end{enumerate}

\section{Results} \label{sec:results}

\subsection{Unbiased sample}

\begin{figure*}
\centering
    \includegraphics[width=0.82\textwidth]{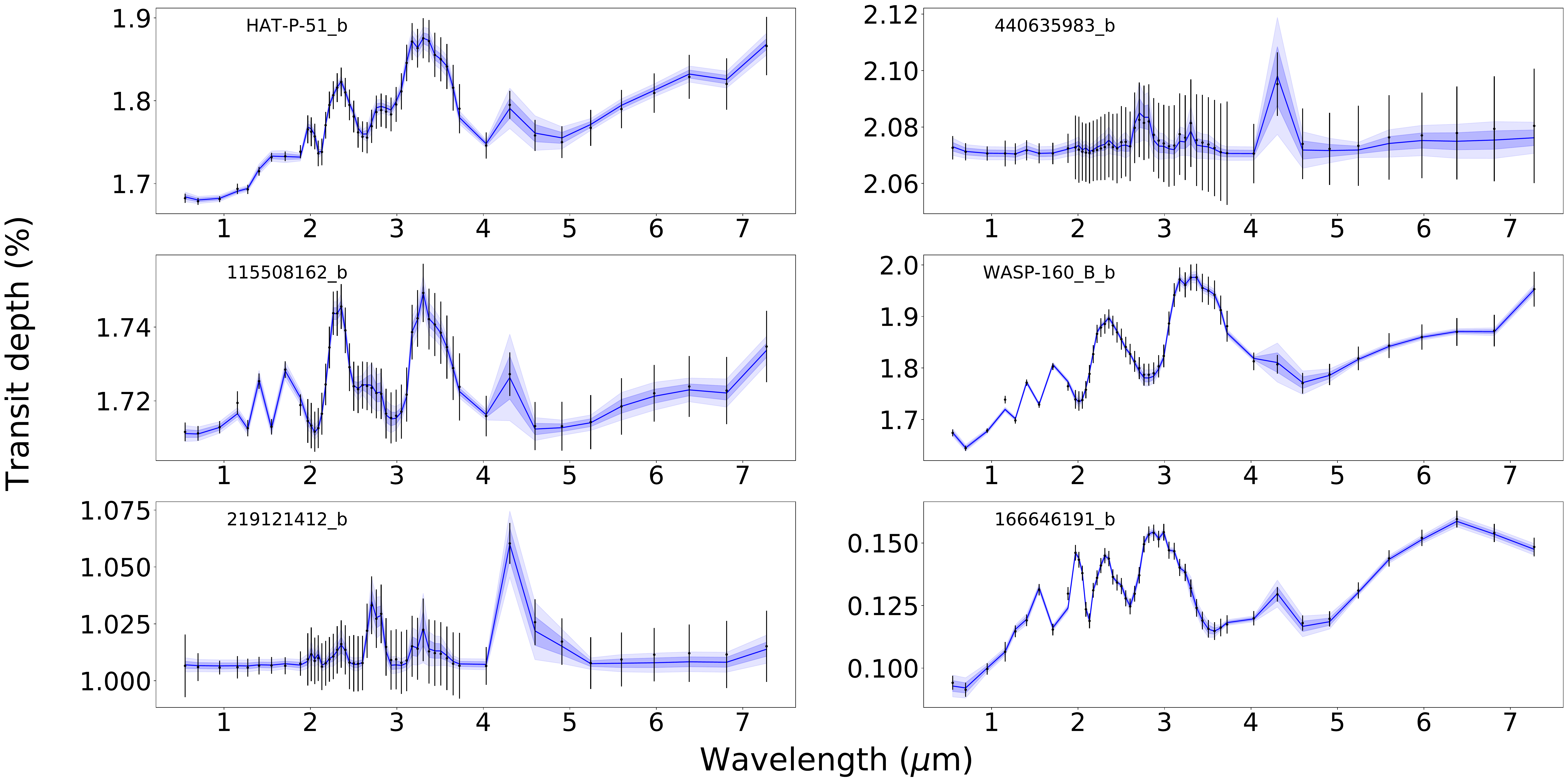}
    \includegraphics[width=0.82\textwidth]{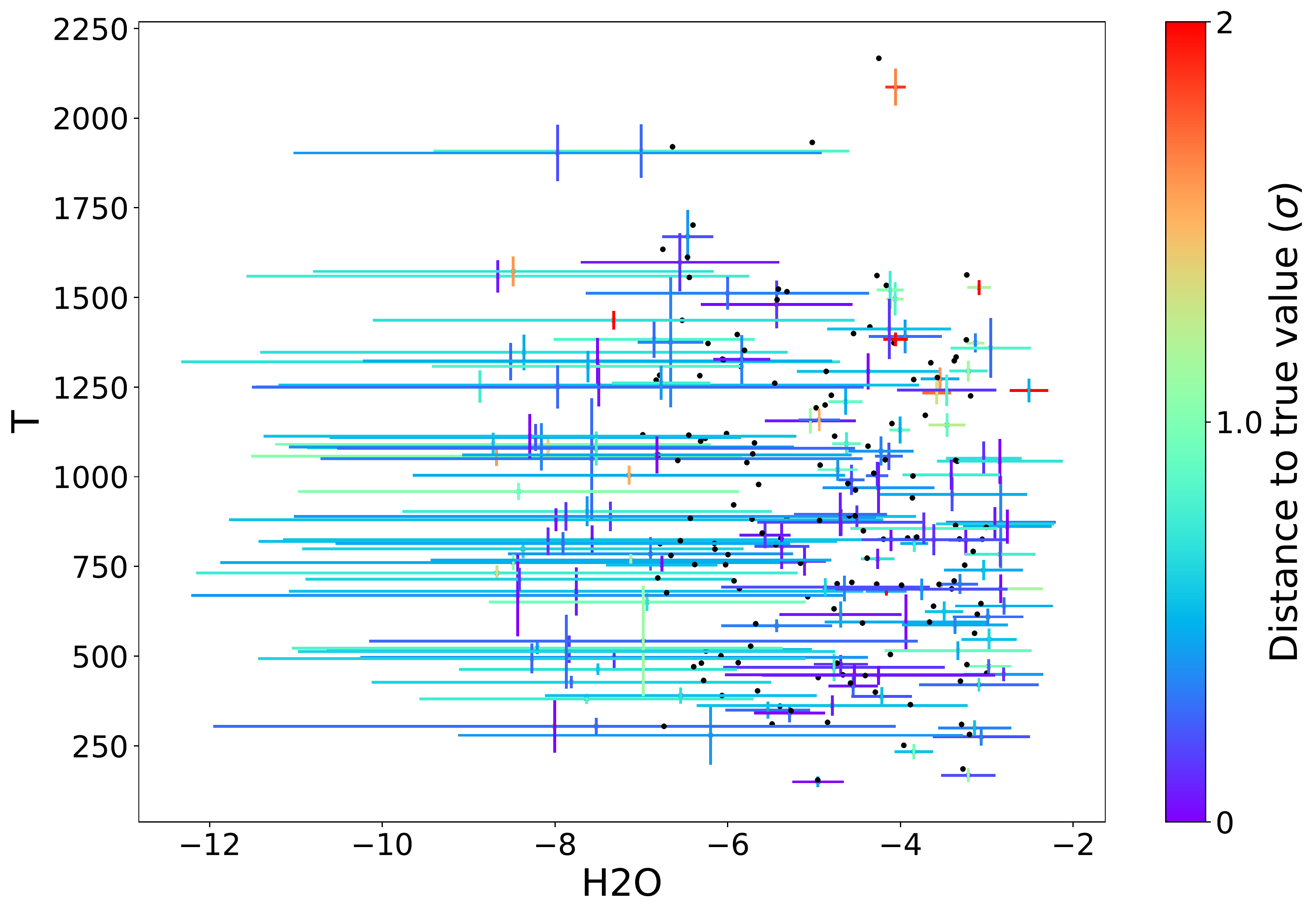}
\caption{Unbiased sample. Top: Observations (black) and best fit spectra (blue) for select planetary atmospheres as observed by Ariel in Tier-2 mode. Bottom: Correlation map between the temperature and the retrieved abundances of water.  We show the retrieved 1-$\sigma$ error bars on the retrieved parameters. The colour-scale represents the distance to the true value (indicated with the black dots) in units of  1-$\sigma$.}
\label{fig:spectra}
\end{figure*}

\begin{figure*}
\centering
    \includegraphics[width=0.73\textwidth]{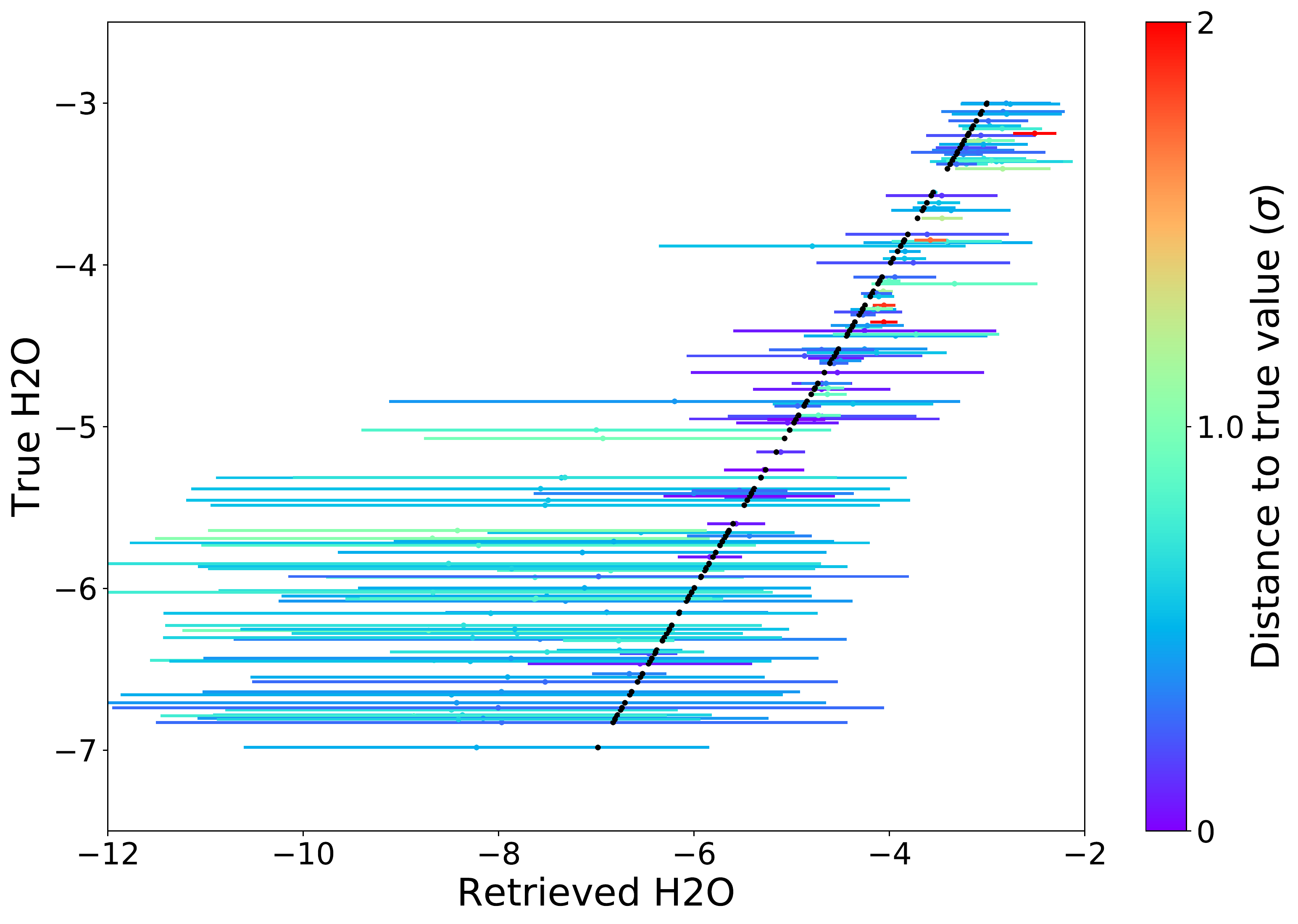}
    \includegraphics[width=0.73\textwidth]{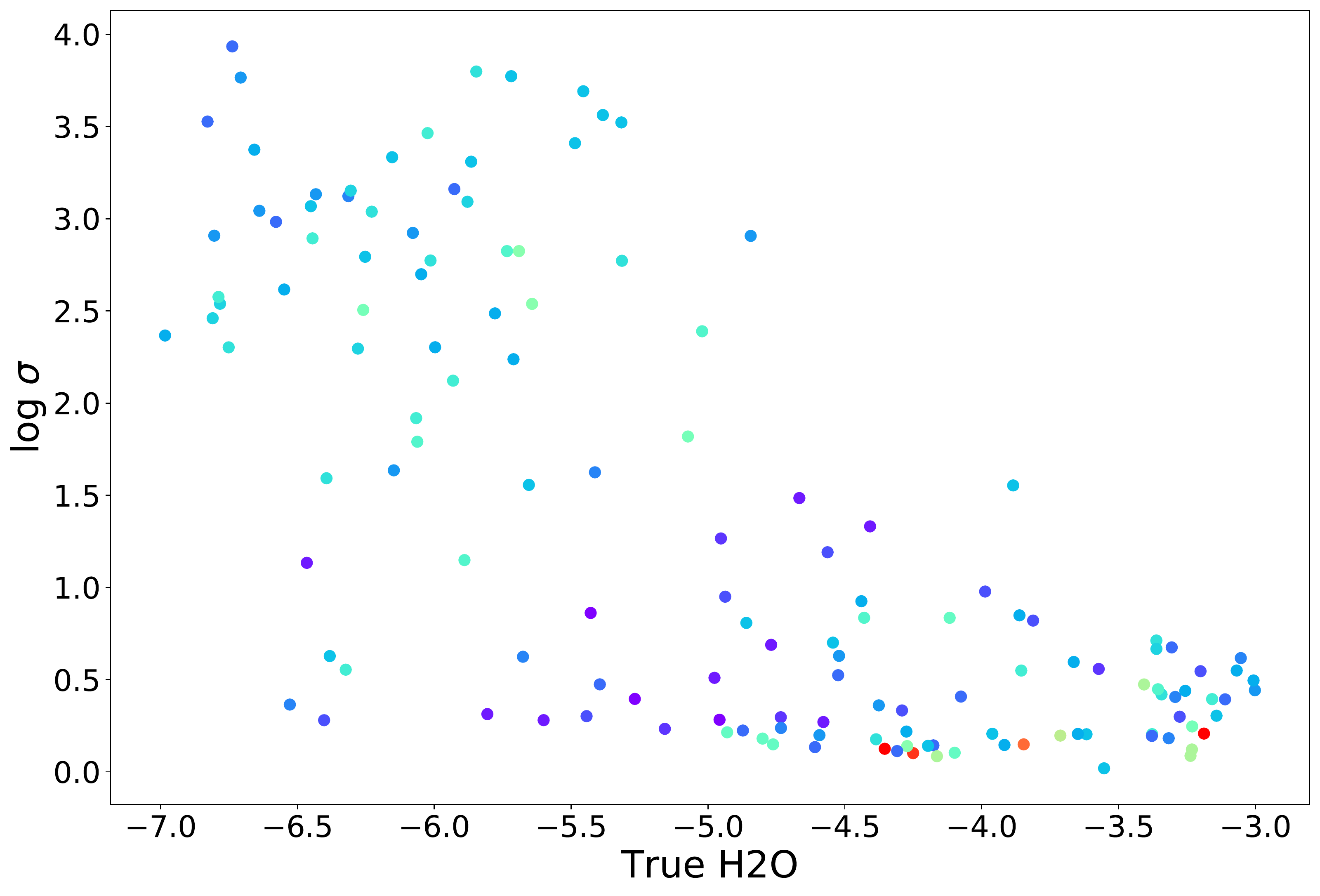}
\caption{Map of the H$_2$O retrieved abundance versus the true value for the unbiased sample. The colour-scale of the 1-$\sigma$ retrieved error bars represents the distance to the true value in units of  1-$\sigma$.}
\label{fig:true_map_h2o}
\end{figure*}

We show  in Figure \ref{fig:spectra} both the observed and retrieved spectra for a subset of the simulated Ariel Tier-2 observations, along with the correlation map between water abundance and temperature with their 1$\sigma$ uncertainties. The distance between the true and the retrieved  value is  visualised by the colour of the point. The retrieved parameters are represented by the median chemical or temperature profiles weighted by the contribution function.  The contribution function is defined as the wavelengths averaged variations of the optical depth with pressure. This choice ensures that the values reported well reflect the conditions in the atmospheric regions probed by observations.  In order to better visualise the Ariel detection limits in Tier 2, we also provide complementary plots of the retrieved abundances versus their true values for each molecule. The H$_2$O map is presented in Figure \ref{fig:true_map_h2o}.

\begin{figure*}[h]
\centerline{\includegraphics[width=0.81\textwidth]{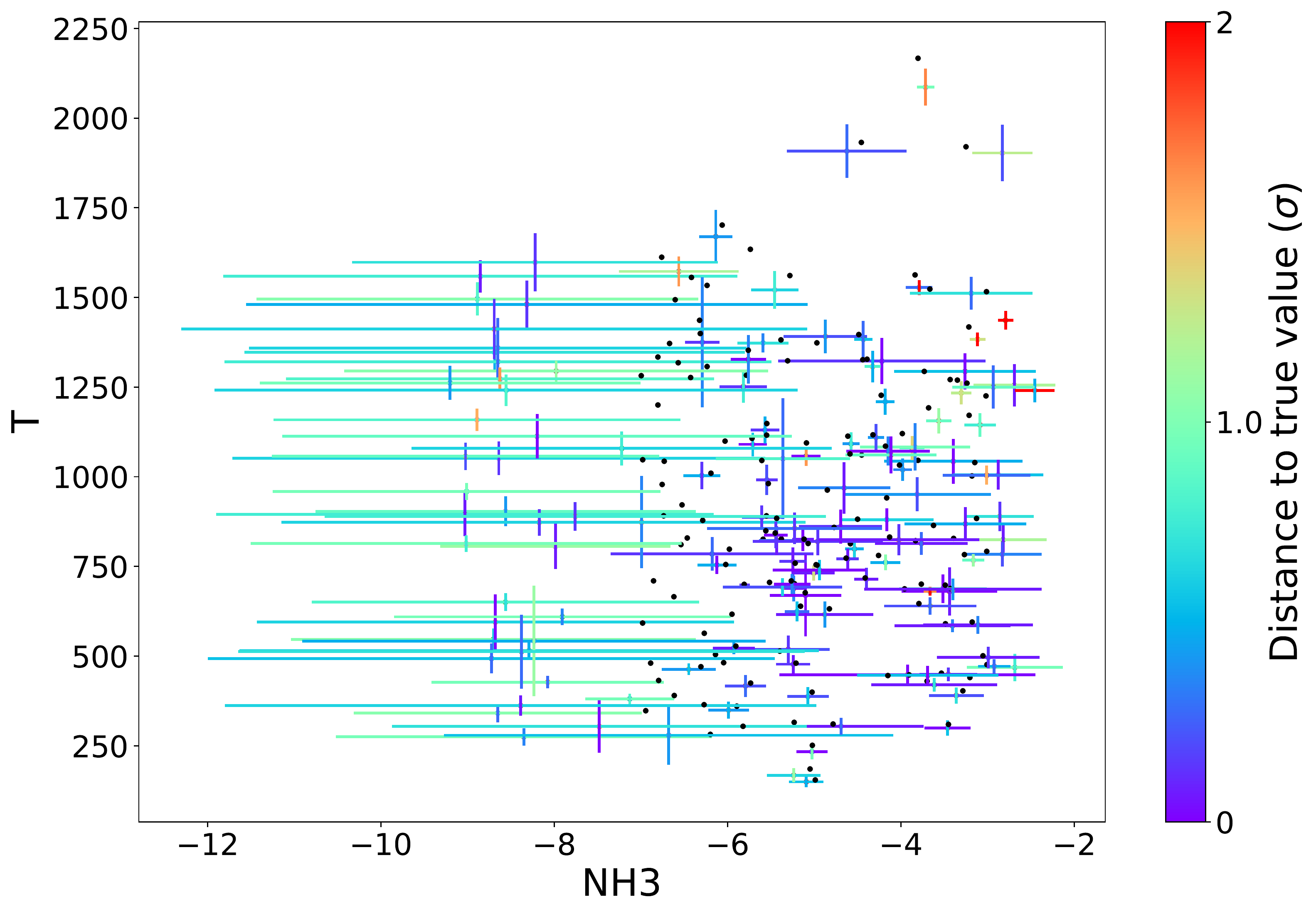}}
\centerline{\includegraphics[width=0.81\textwidth]{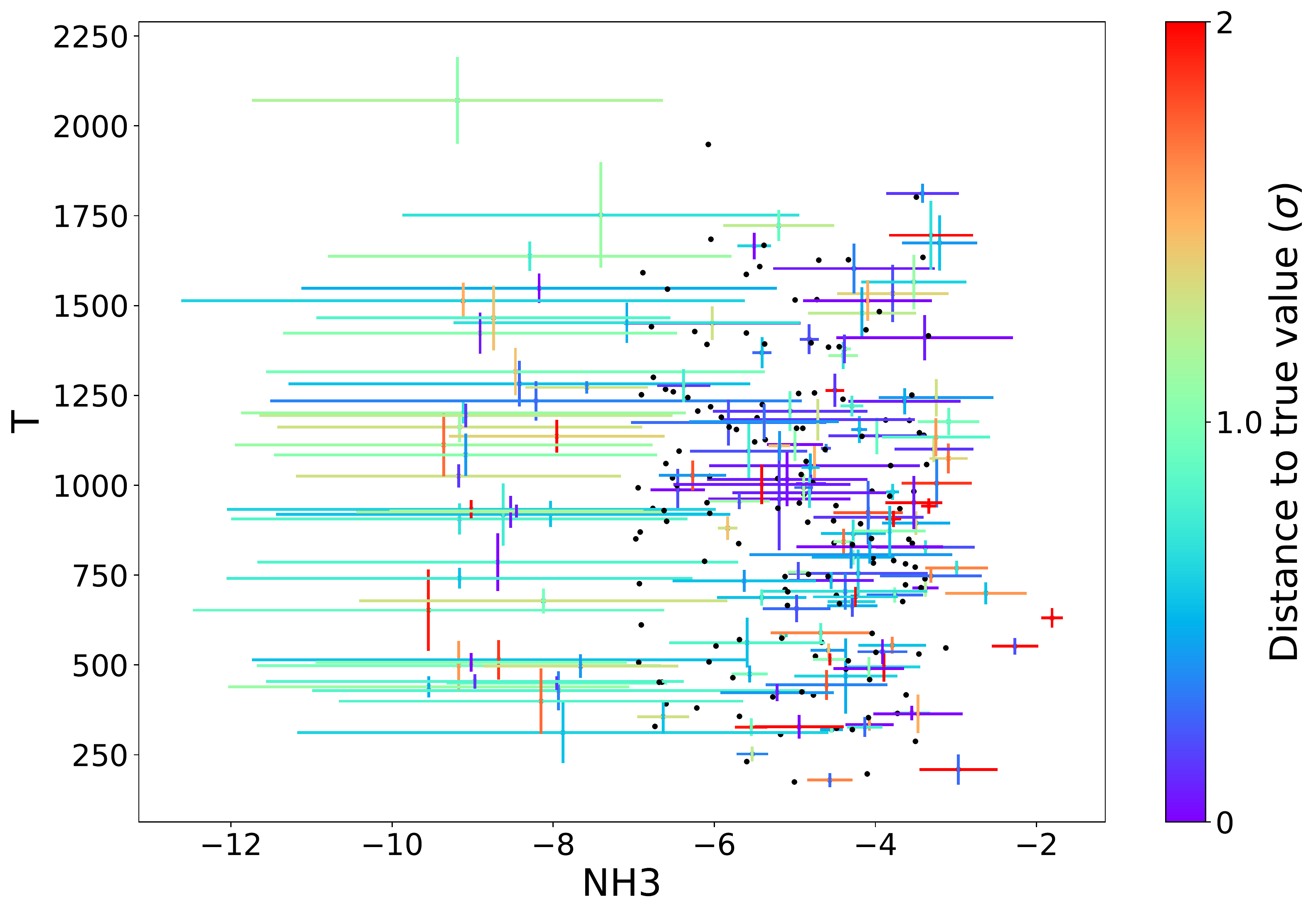}}
\caption{Unbiased sample: correlation map between the temperature and the retrieved abundances of NH$_3$, with the 1-$\sigma$ retrieved error bars. The colour-scale represents the distance to the true value in units of  1-$\sigma$. Top: Non-scattered spectra. Bottom: Scattered spectra.}
\label{fig:sc_nh3}
\end{figure*}

\begin{figure*}[h]
\centerline{\includegraphics[width=0.81\textwidth]{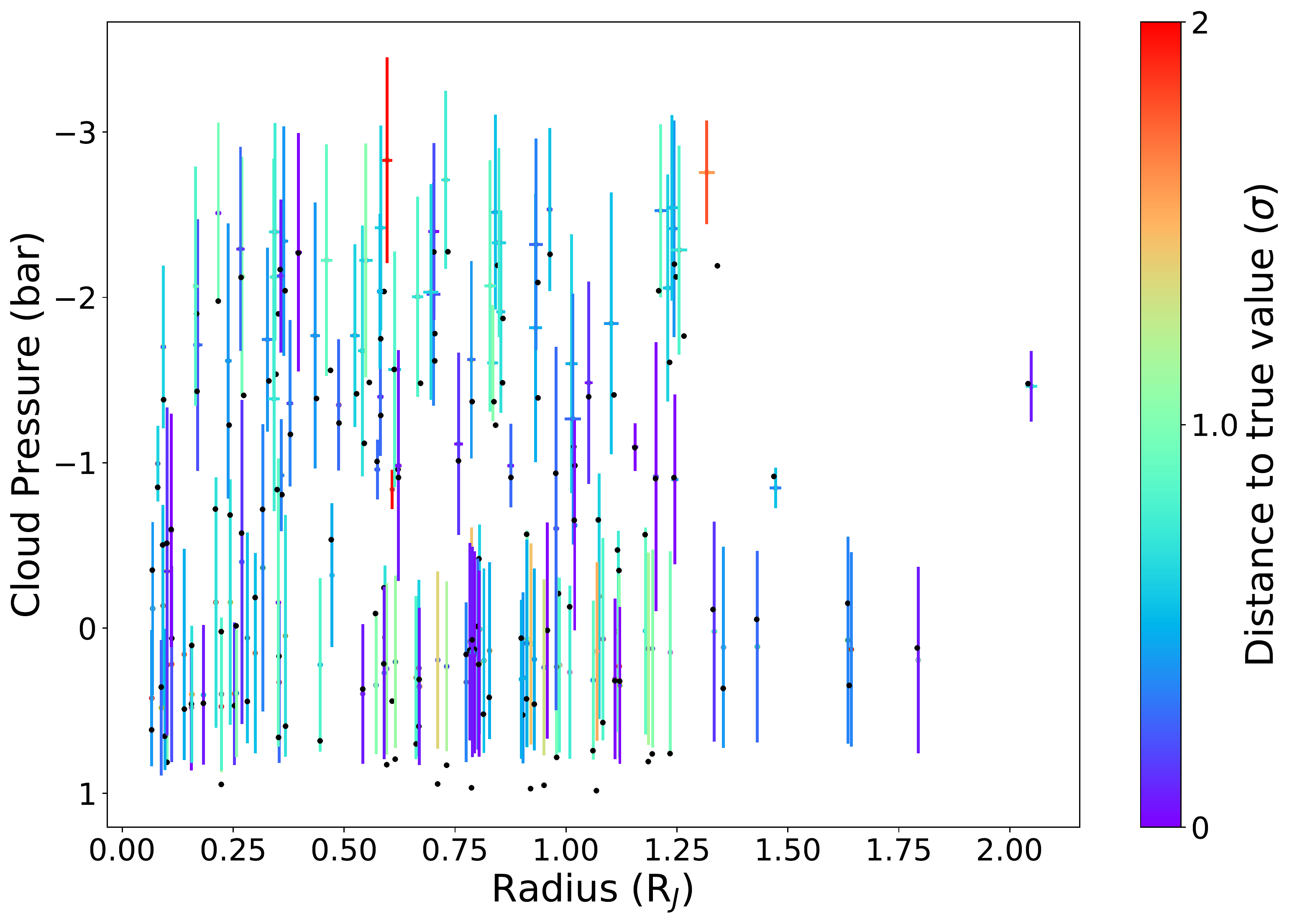}}
\centerline{\includegraphics[width=0.81\textwidth]{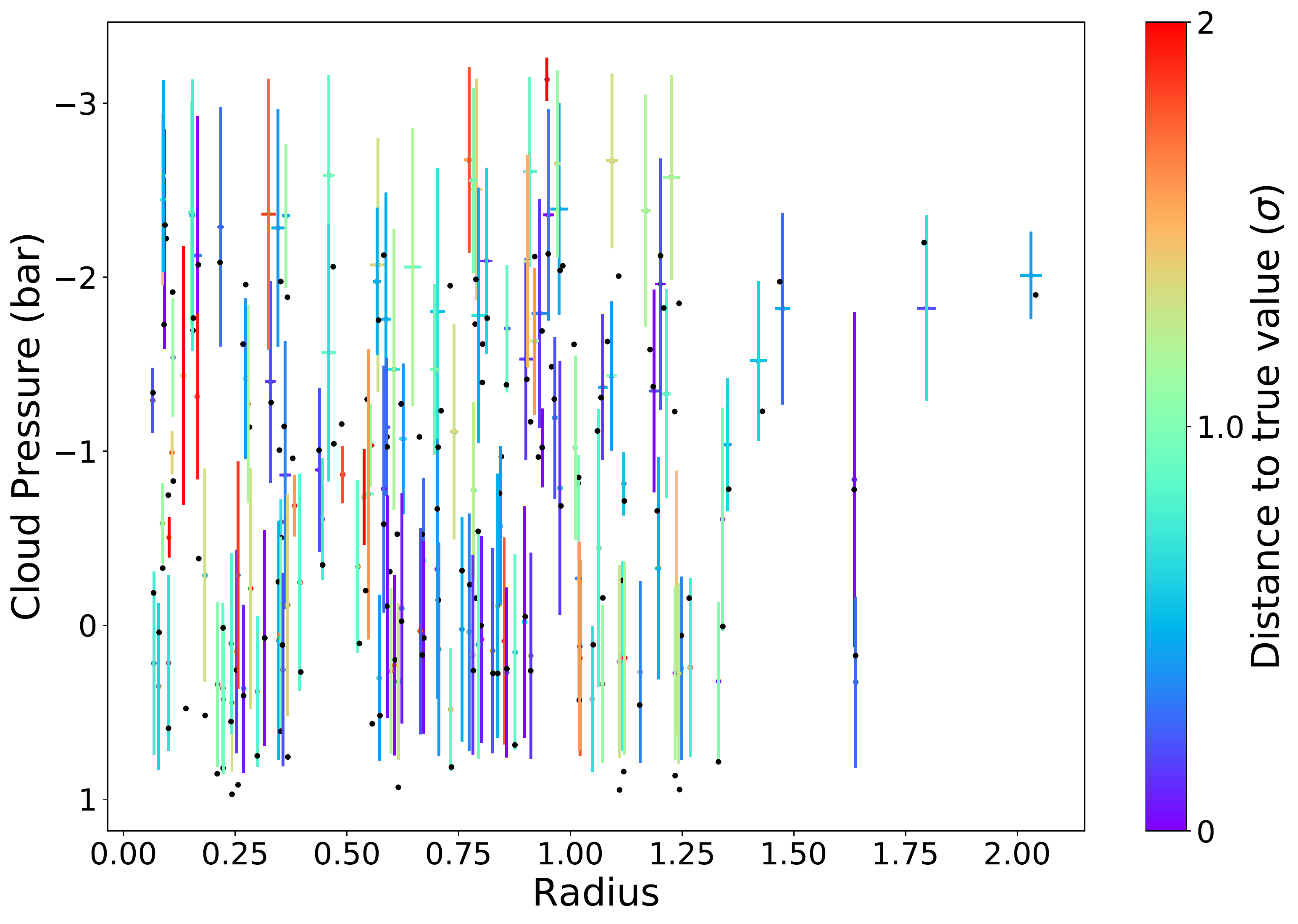}}
\caption{Unbiased sample: correlation map between the clouds and the radius, with the 1-$\sigma$ retrieved error bars. The colour-scale represents the distance to the true value in units of  1-$\sigma$. Top: Non-scattered spectra. Bottom: Scattered spectra.}
\label{fig:clouds}
\end{figure*}

The water-temperature map  in   Figure \ref{fig:spectra} clearly shows that our unbiased population is randomly spread in the parameter space, as expected. The retrieved temperature is very precise across the whole parameter space, showcasing the ability of Ariel Tier-2 to study a wide range of planets.
It also illustrates that the retrieved values are mostly accurate  for water abundances higher than $10^{-6}$: with the exception of a few cases, the retrieved values for water and temperature fall well within the 1$\sigma$ error bars (blue to green in the colour scale).  We notice for water a rapid change in the posteriors for abundances smaller than  $10^{-5}$, marked by large error bars on the left side of the plot. Indeed, when the abundance is too low, the retrievals are not able to distinguish well the features and  provide only upper limits. This is an expected behaviour and an indication of the Ariel detection limit for our sample of planets. This exercise was repeated for  other molecules to assess Ariel ability  to detect different sets of molecules in Tier-2 mode. Other temperature-molecule maps, as well as the radius-clouds map, are reported in  Appendix (Figures \ref{fig:sc_ch4}, \ref{fig:sc_co}, \ref{fig:sc_co2}, \ref{fig:sc_nh3}, \ref{fig:clouds}).

\begin{figure*}
\centering
    \includegraphics[width=0.82\textwidth]{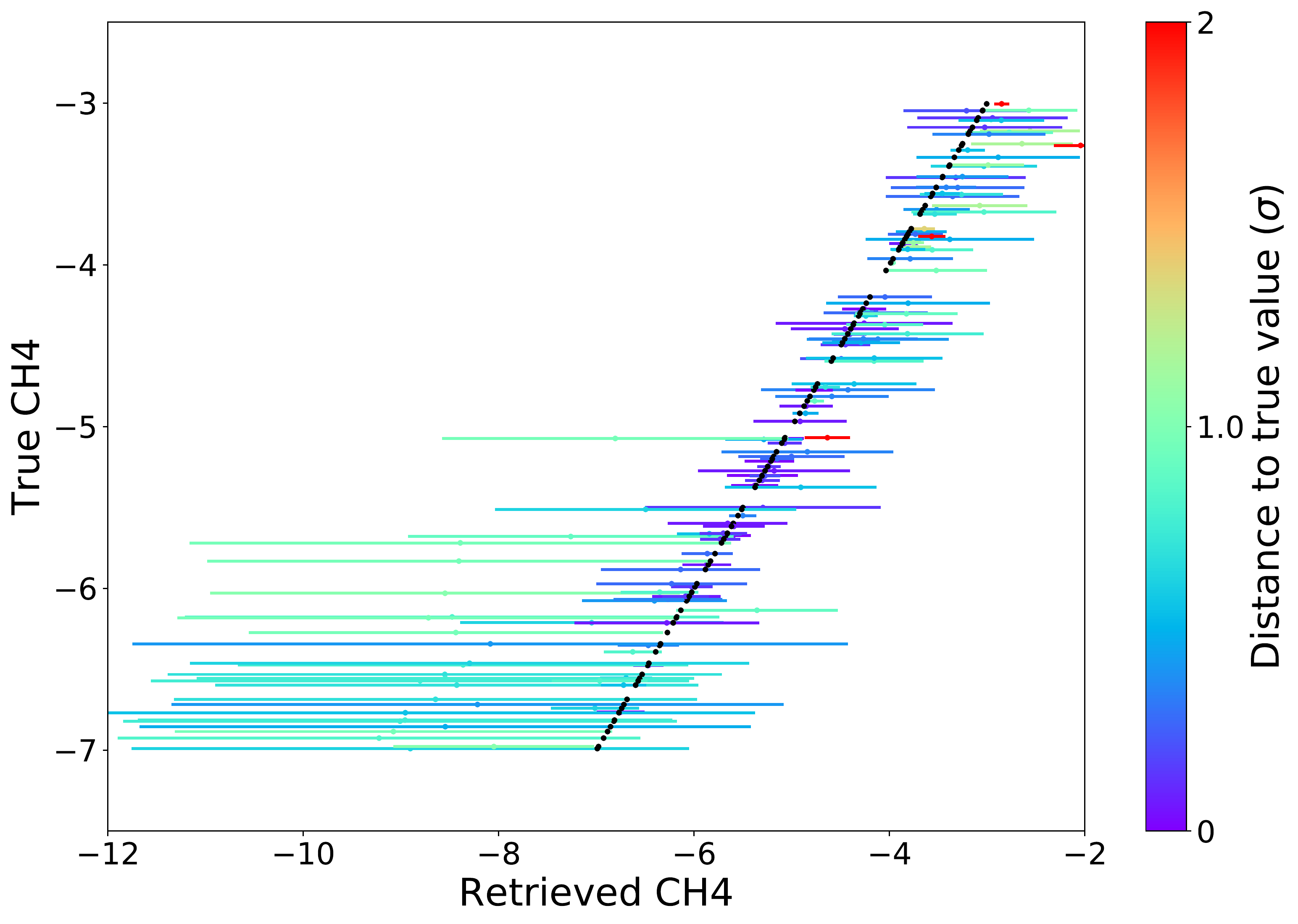}
    \includegraphics[width=0.82\textwidth]{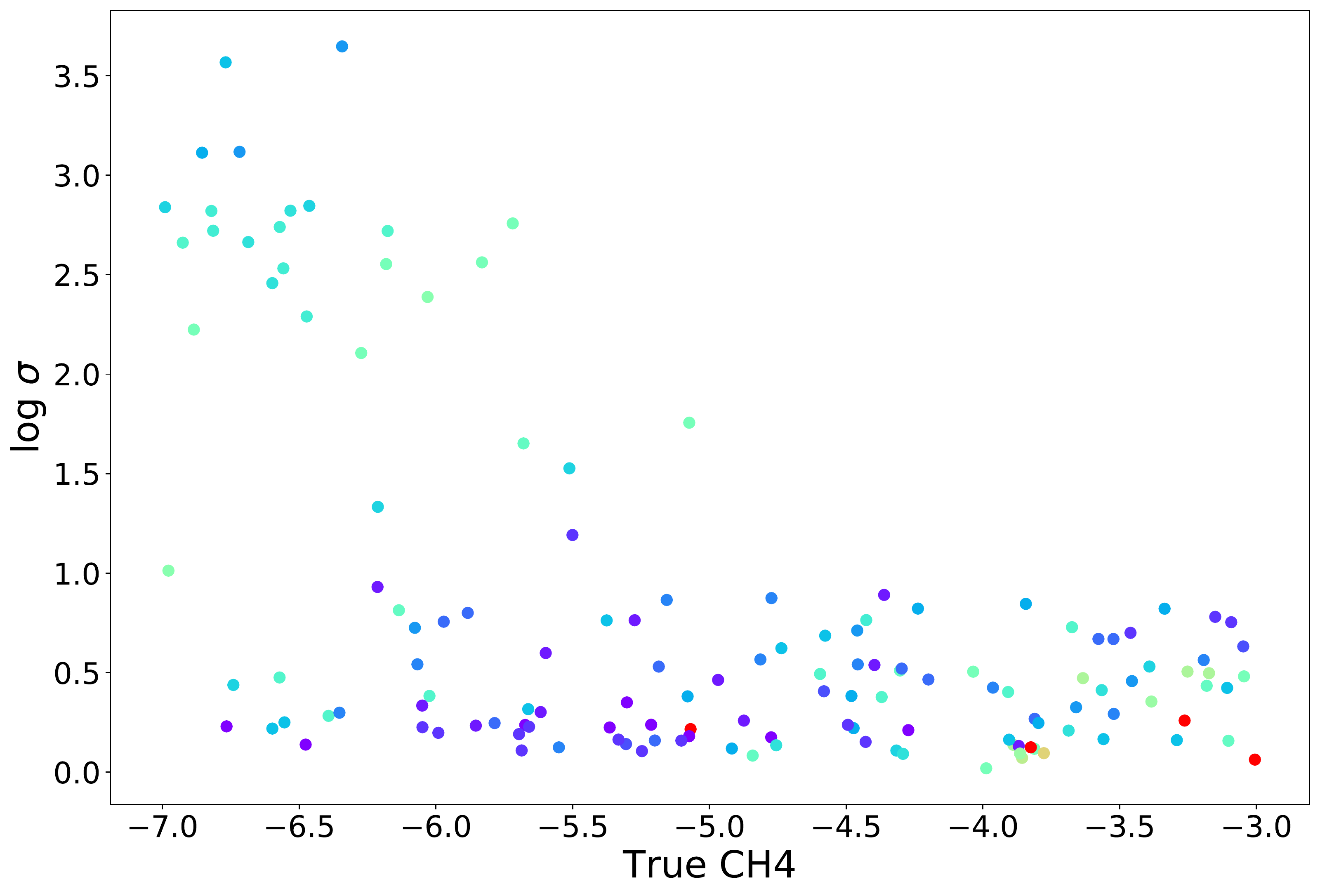}
\caption{Top: Map of the CH$_4$ retrieved abundances versus their the values for the unbiased sample. Bottom: Error retrieved as a function of the input abundances. The colour-scale of the 1-$\sigma$ retrieved error bars represents the distance to the true value in units of  1-$\sigma$.}
\label{fig:true_map_ch4}
\end{figure*}

\begin{figure*}
\centering
    \includegraphics[width=0.82\textwidth]{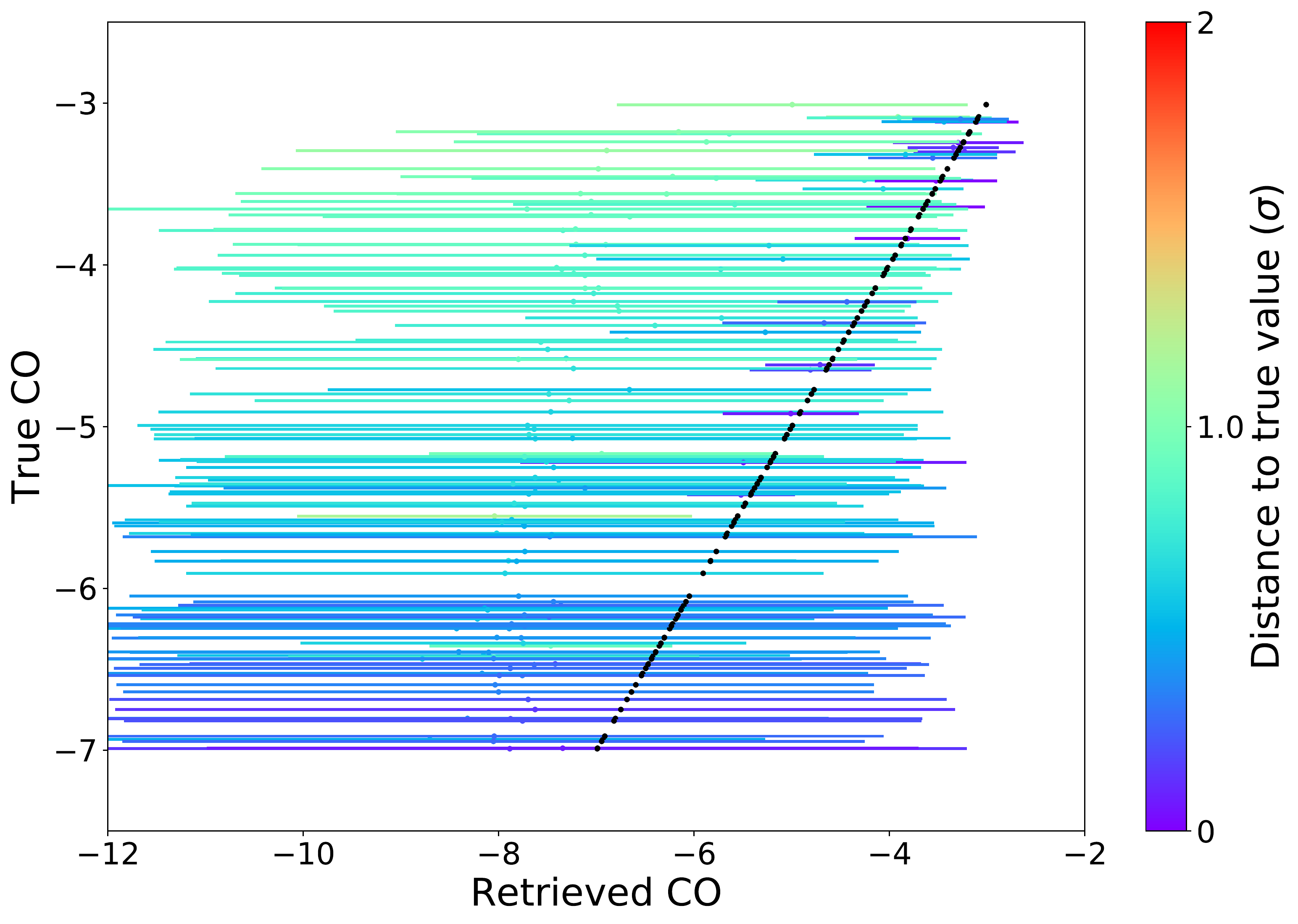}
    \includegraphics[width=0.82\textwidth]{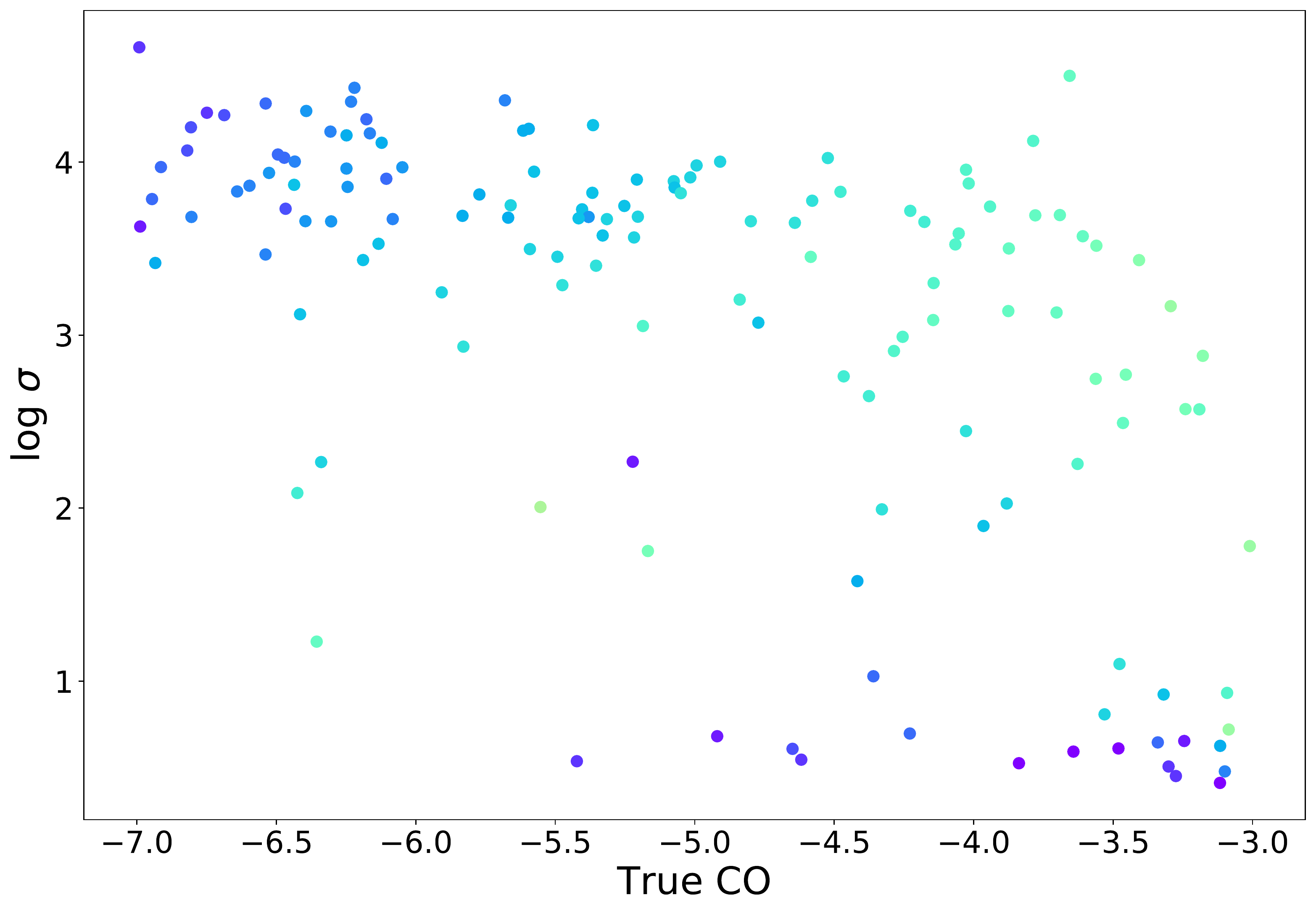}
\caption{Top: Map of the CO retrieved abundances versus their the values for the unbiased sample. Bottom: Error retrieved as a function of the input abundances. The colour-scale of the 1-$\sigma$ retrieved error bars represents the distance to the true value in units of  1-$\sigma$.}
\label{fig:true_map_co}
\end{figure*}

\begin{figure*}
\centering
    \includegraphics[width=0.82\textwidth]{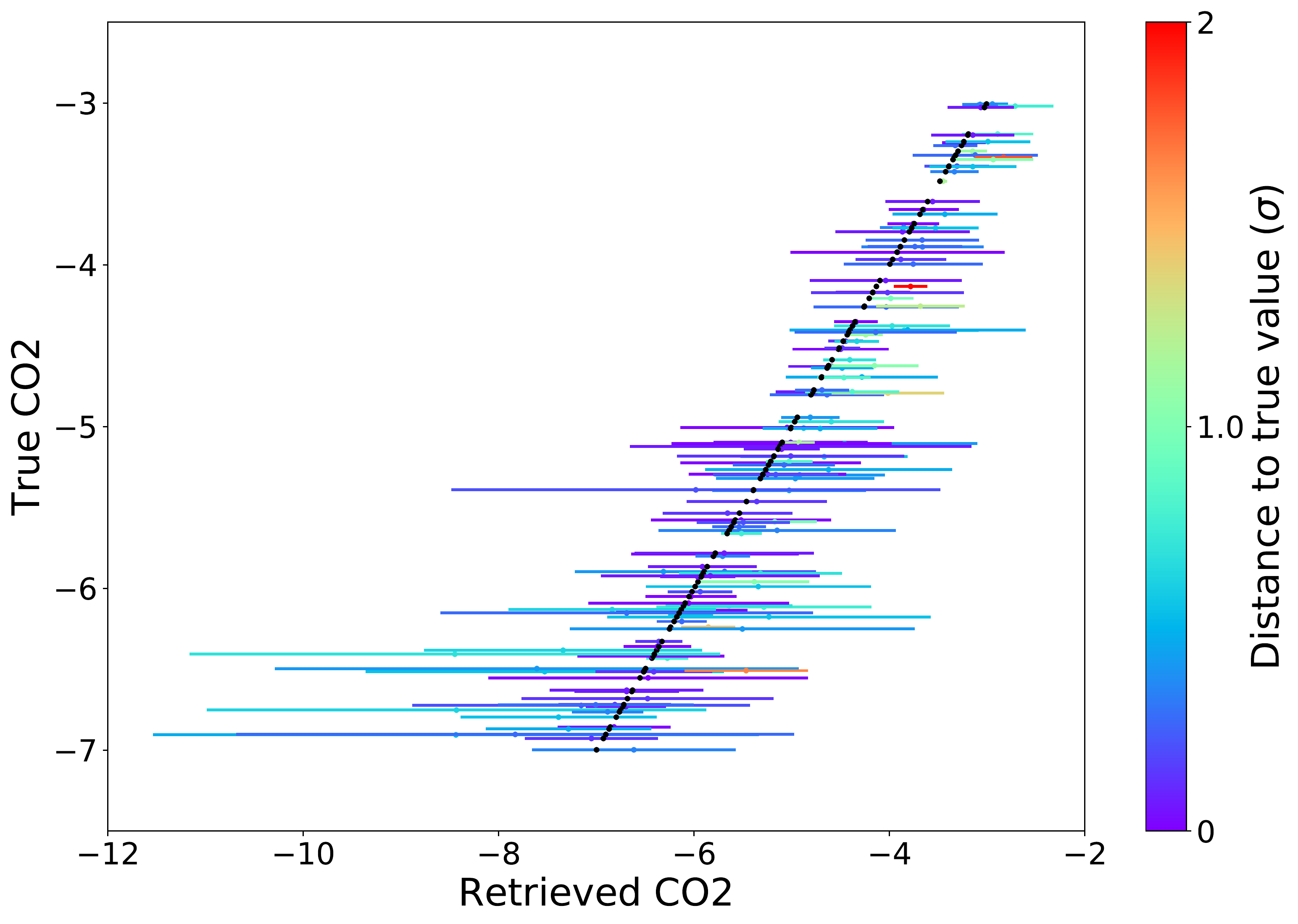}
    \includegraphics[width=0.82\textwidth]{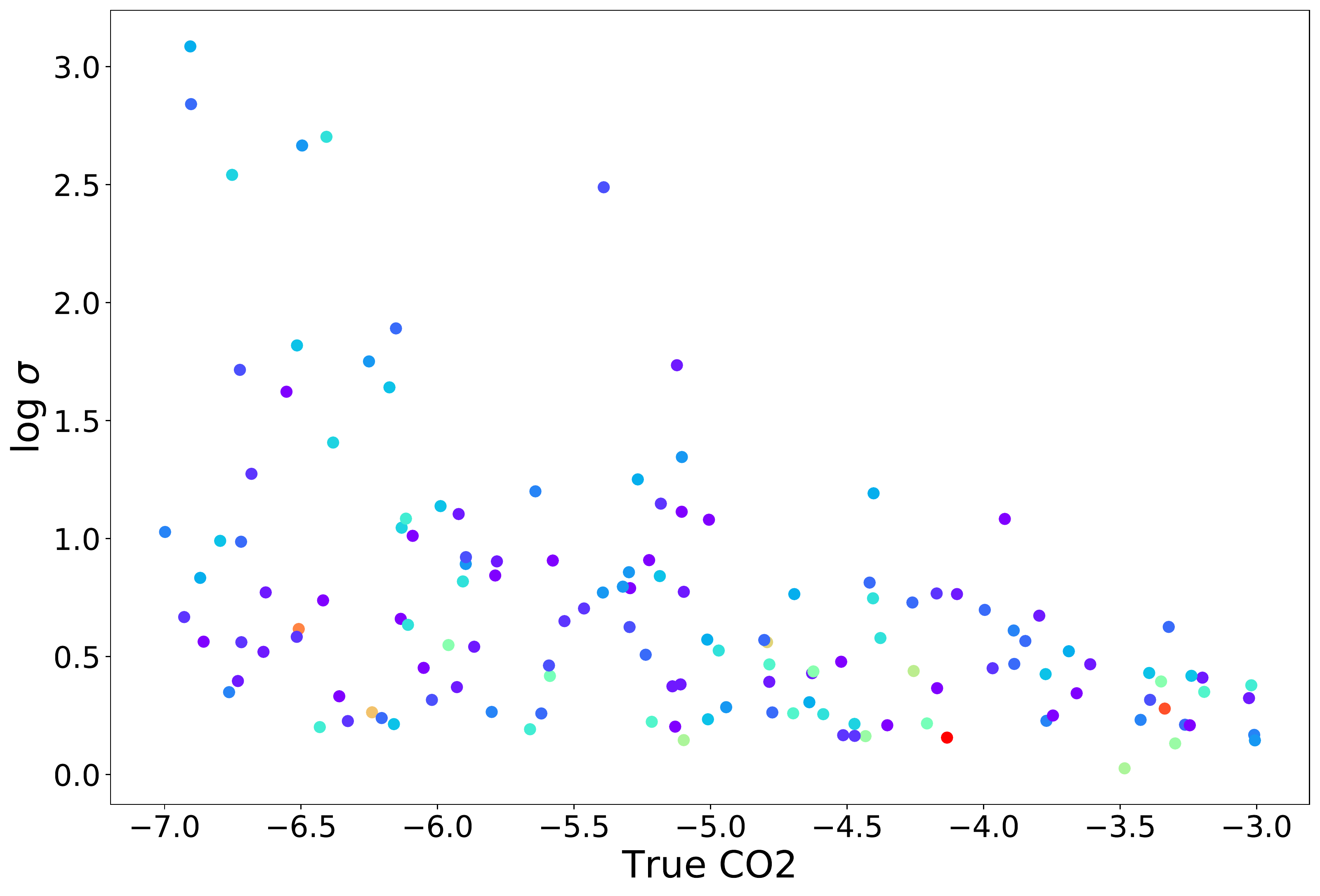}
\caption{Top: Map of the CO$_2$ retrieved abundances versus their the values for the unbiased sample. Bottom: Error retrieved as a function of the input abundances. The colour-scale of the 1-$\sigma$ retrieved error bars represents the distance to the true value in units of  1-$\sigma$.}
\label{fig:true_map_co2}
\end{figure*}

\begin{figure*}
\centering
    \includegraphics[width=0.82\textwidth]{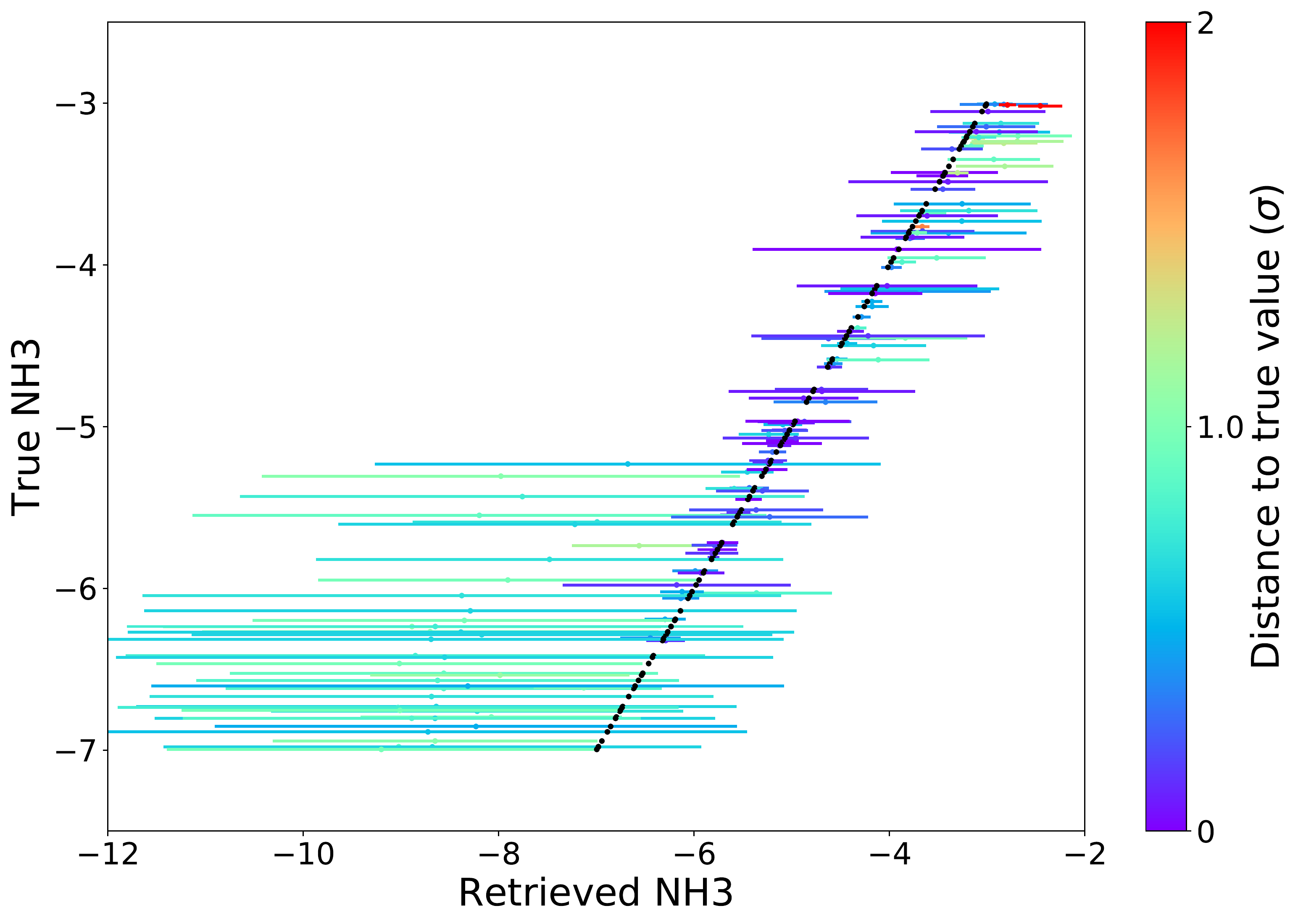}
    \includegraphics[width=0.82\textwidth]{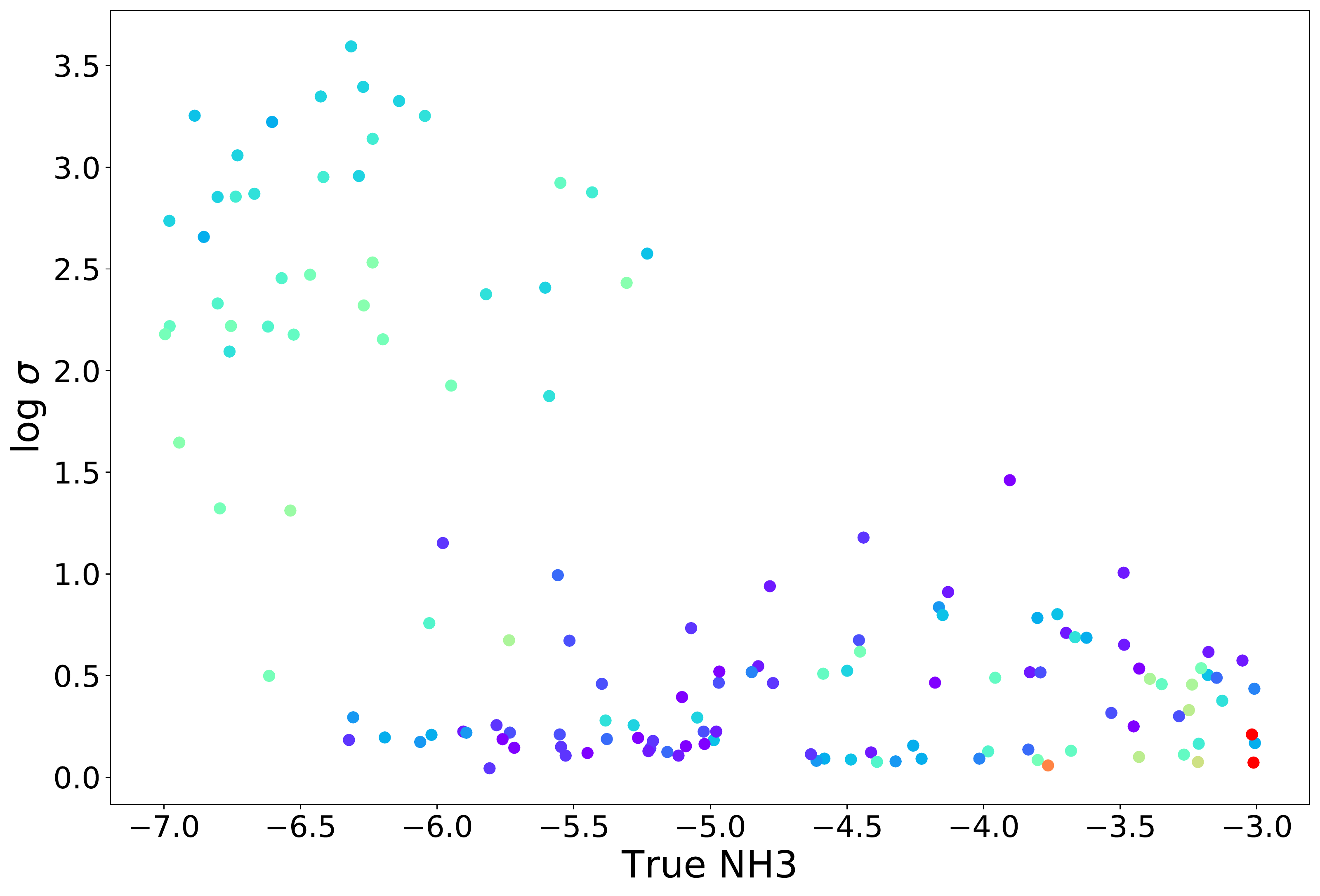}
\caption{Top: Map of the NH$_3$ retrieved abundances versus their the values for the unbiased sample. Bottom: Error retrieved as a function of the input abundances. The colour-scale of the 1-$\sigma$ retrieved error bars represents the distance to the true value in units of  1-$\sigma$.}
\label{fig:true_map_nh3}
\end{figure*}

The detection limits are best visualised in the retrieved versus true abundances (see Figure \ref{fig:true_map_h2o}). In the same Figure, we also show the retrieved uncertainties versus input abundances as this allows us to distinguish 3 regimes. The first regime corresponds to low abundances where molecular detections are not possible: for example, between 10$^{-7}$ and 10$^{-6}$ for water no detections seem possible with Ariel. Other molecules are presented in Figures \ref{fig:true_map_ch4}: CH$_4$; Figure \ref{fig:true_map_co}: CO; Figure \ref{fig:true_map_co2}: CO$_2$ and Figure \ref{fig:true_map_nh3}: NH$_3$. It is interesting to note that when the molecules are not detected, the retrieved errors ($\sigma$) are dominated by the size of the priors and the location of the detection limit: for water no-detection errors are between 2 and 4 orders of magnitude. The second regime for intermediate abundances displays a mix between successful detections and lack of evidence for the molecules. This corresponds to the region with large ranges in the retrieved errors (between 10$^{-6}$ and 10$^{-5}$ for water). In general, this variability is due to the other constituents in the planet that are susceptible to mask the signal of interest (e.g: clouds, other molecules). Finally, for the highest abundances, the retrieved uncertainties are low (less than 1 order of magnitude in the mixing ratios), which indicate that these abundances are always retrieved, regardless of the other constituents in the atmosphere. 

Additionally, the map exploring the correlation between planetary radius and cloud top pressure shows that Ariel can separate well these parameters, most likely thanks to the FGS optical channels. 

\begin{table}
\centering

$\begin{array}{|c|c|c|}
\mbox{Molecule} & \mbox{Tier 2} & \mbox{Tier 3} \\
\hline
\mbox{log(H$_2$O)} & \mbox{$-6.5$} & \mbox{$< -7$} \\
\mbox{log(CH$_4$)} & \mbox{$-7$} & \mbox{$< -7$} \\
\mbox{log(CO)} & \mbox{$-5.5$} & \mbox{$-6$} \\
\mbox{log(CO$_2$)} & \mbox{$-7$} & \mbox{$< -7$} \\
\mbox{log(NH$_3$)} & \mbox{$-6.5$} & \mbox{$< -7$} \\
\end{array}$
\caption{Detection limits for each molecule in Ariel Tier-2 and Tier-3 samples considered here. The detection limits corresponds to the lowest value we were able to extract abundance with less than 1 order of magnitude errors. Tier-3 sample includes only 14  planets.}
\label{tab:detection_limit}
\end{table}

We repeated the same experiment with the second run composed of 'scattered' spectra. Each planet is simulated with a new set of randomised parameters. As previously stated, the observed values of the transit depth are assumed to follow a normal distribution (the mean is the simulated transit depth and the standard deviation is the instrumental noise), which better reproduces a real observation. Figure \ref{fig:spectra_scattered} shows the  water-temperature map. The other chemical parameters are reported in  Appendix (Figures \ref{fig:sc_ch4}, \ref{fig:sc_co}, \ref{fig:sc_co2}, \ref{fig:sc_nh3}, \ref{fig:clouds}). 
From the analysis of the scattered spectra, we  appreciate that the scattering of the data points around their true value does not necessary introduce biases in Ariel Tier-2 retrieval studies. Indeed, this result, which has already been explored in \cite{Feng_retrieval_earthanalog, Changeat_2019_2layer}, naturally arises from the redundancy of the information relative to each molecule in the Ariel spectra and the fact that in most cases N repeated observations are needed to obtain Tier-2 requirements, therefore reducing by $1/\sqrt{N}$ the scattering amplitude around their true value.  \cite{Feng_retrieval_earthanalog} highlighted that,  to  avoid potential biases arising from individual noise instances, one would essentially have to produce multiple retrievals with different noise instances and average the obtained results. As this was not computationally feasible, they chose  not to scatter the spectra and use the true value as an approximation, stating that the shape of the posteriors would be accurate but that the position may be optimistically centred. For  Ariel Tier-2 observations, the information content of the spectra is redundant enough to ensure that the retrieved values are not affected by this phenomenon and these are mostly centred around the true value in both scattered and non-scattered scenarios. For all molecules, we find that the correlation maps are very similar in both cases and the detection limits remain unchanged from the non-scattered runs. For the clouds, however, we note an overall increase in the distance to the true value (see correlation map in Appendix Figure \ref{fig:clouds}).  We note that in the simulations presented here, we considered  fully opaque grey cloud cover, which is essentially the worst case scenario as no cloud features are detectable and it is well known to be degenerate with radius \citep{changeat_2020_mass}. More realistic cloud simulations will be considered in a future paper to test more thoroughly this case.

\begin{figure*}
\centering
\centerline{\includegraphics[width=0.82\textwidth]{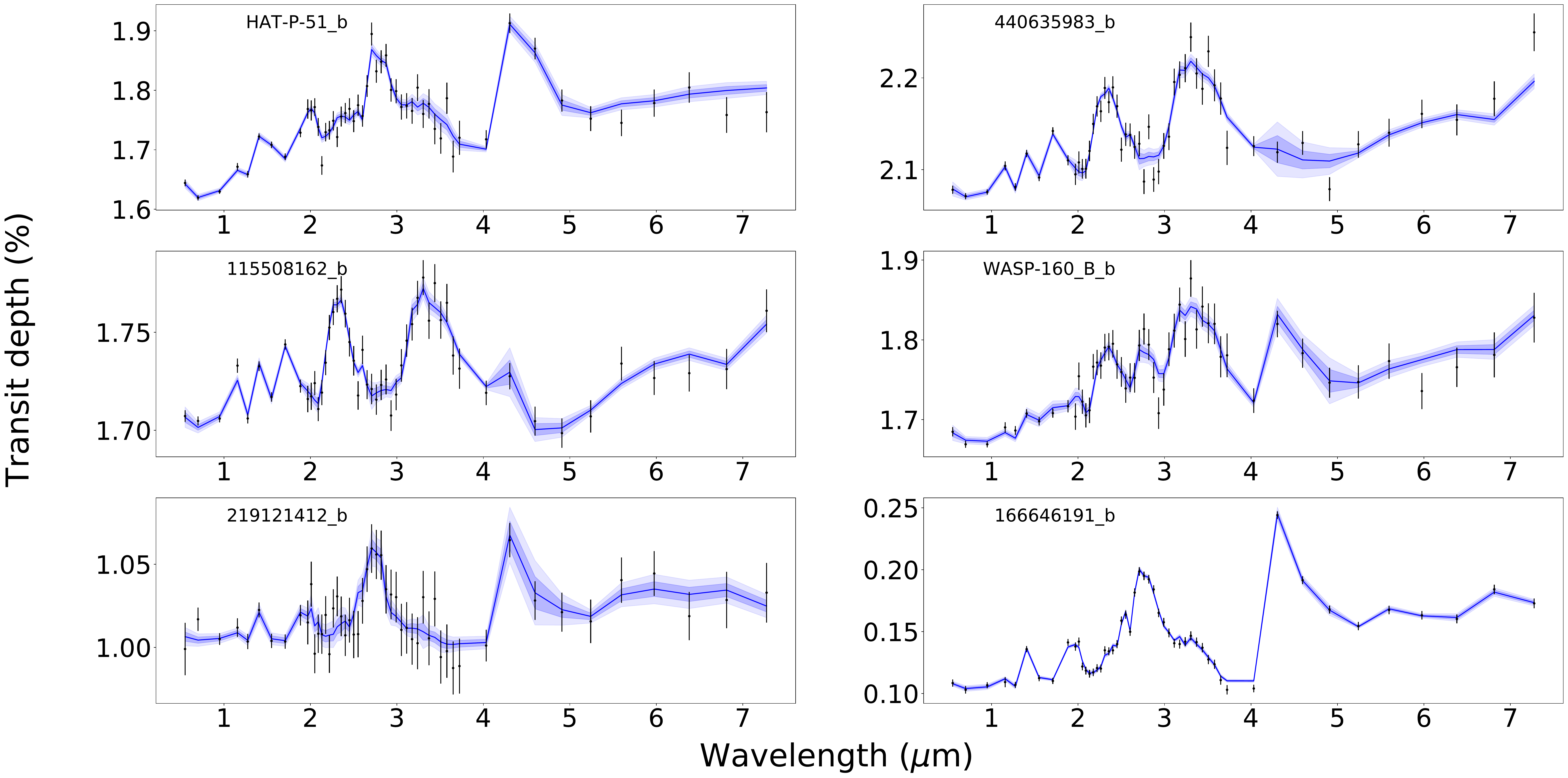}}
\centerline{\includegraphics[width=0.82\textwidth]{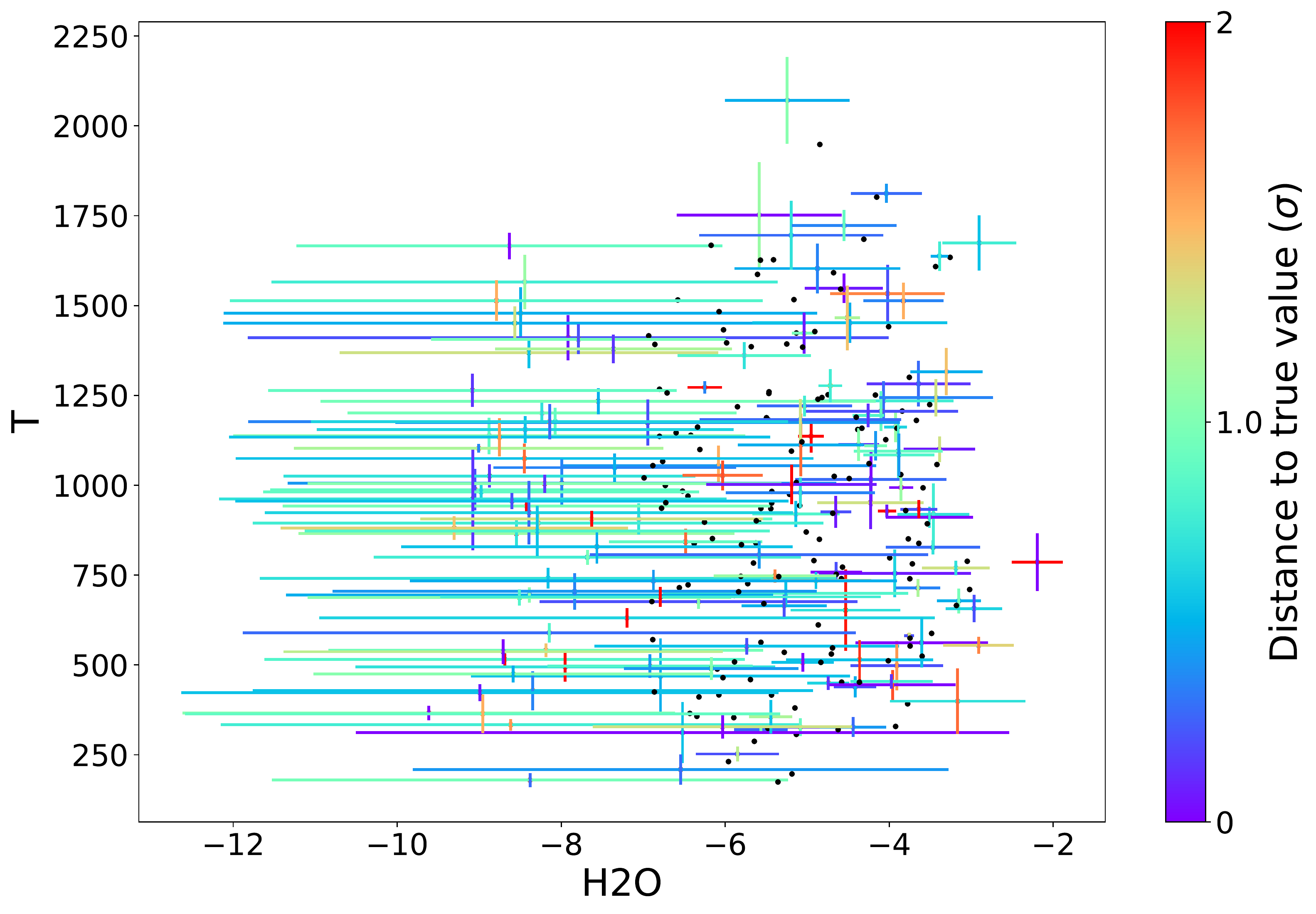}}
\caption{ Same as Figure \ref{fig:spectra} but in this new run, we scatter the Ariel spectra around their true values. }
\label{fig:spectra_scattered}
\end{figure*}

\begin{figure*}[h]
\centering
\includegraphics[width=0.98\textwidth]{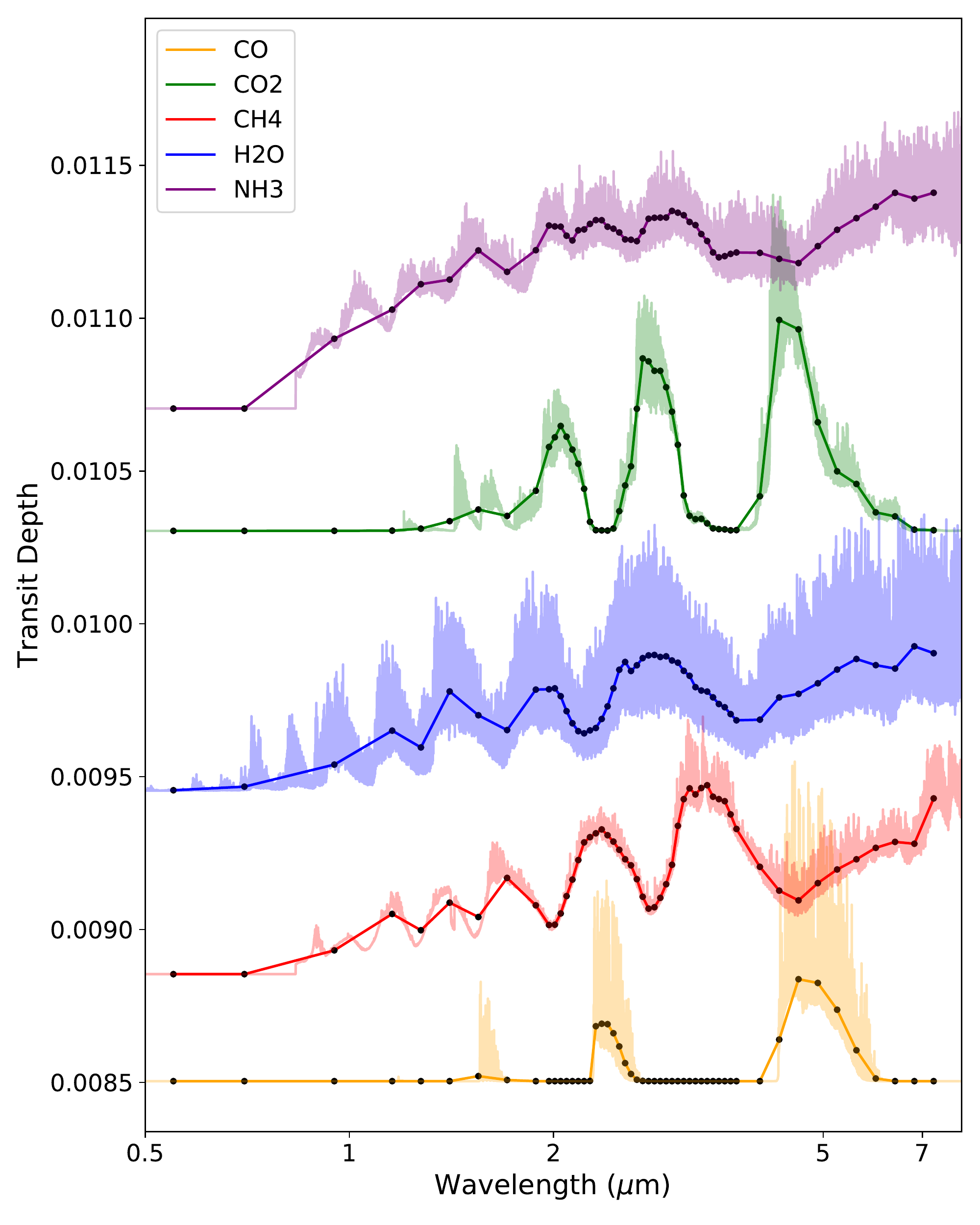}

\caption{Absorption in Ariel of the molecular species considered in this paper. Each simulation is for a 1 R$_J$, 1 M$_J$ planet and 1 R$_S$ star with 10$^{-5}$ of the considered molecule as only absorber. The models are offset for better visibility. The shaded region is the full resolution contribution, while the solid lines and black points correspond to Ariel resolutions. 
}
\label{fig:contrib}
\end{figure*}

We summarise  in Table \ref{tab:detection_limit} the approximate detection limits for each molecule considered. These represent the regions where our retrieval analysis have been able to extract constraints on the given atmospheric constituents. The stated detection limit corresponds to the lowest value that was successfully recovered with less than 1 order of magnitude uncertainties. As seen before, 3 regions of the parameter space can be identified: region 1 with no possible detections of the molecule, region 2 with detections depending on the other atmospheric properties and region 3 where the molecule is always detected. This means the stated values do not represent a guaranty of detection, but rather the lowest limit we can hope to detect the molecule. In addition, we plot the contribution of each molecule individually in Figure \ref{fig:contrib} (each spectrum only contains 10$^{-5}$ of the considered molecule) to show the features span by each molecule. 
In general, Ariel Tier-2 spectra should enable molecular detections down to mixing ratios of $10^{-6}$. In our simulations, only CO appears to be difficult to detect at abundances smaller than $10^{-4}$.  CO presents two features that are overlapping with CO$_2$ at 4.5$\mu$m and with CH$_4$ at 2.5$\mu$m and are relatively weak. In a real scenario (equilibrium chemistry), we believe CO could be more easily distinguishable as our unbiased assumption underestimates the CO abundance and overestimates the CO$_2$ abundance by design \citep{Agundez_2012_eqchem, venot_chem_HJ, Venot_chem_model}. We also note that H$_2$O and CH$_4$ have a large number of anti-correlated features, which may give rise to more featureless spectra when the two molecules are present. For all parameters, Ariel Tier-2 spectra provide accurate and precise estimates, as most of the retrieved error bars are less than 1-sigma away from the true value. This statement applies to both non-scattered and scattered spectra.

\begin{figure*}[h]
\centering
\centerline{\includegraphics[width=0.81\textwidth]{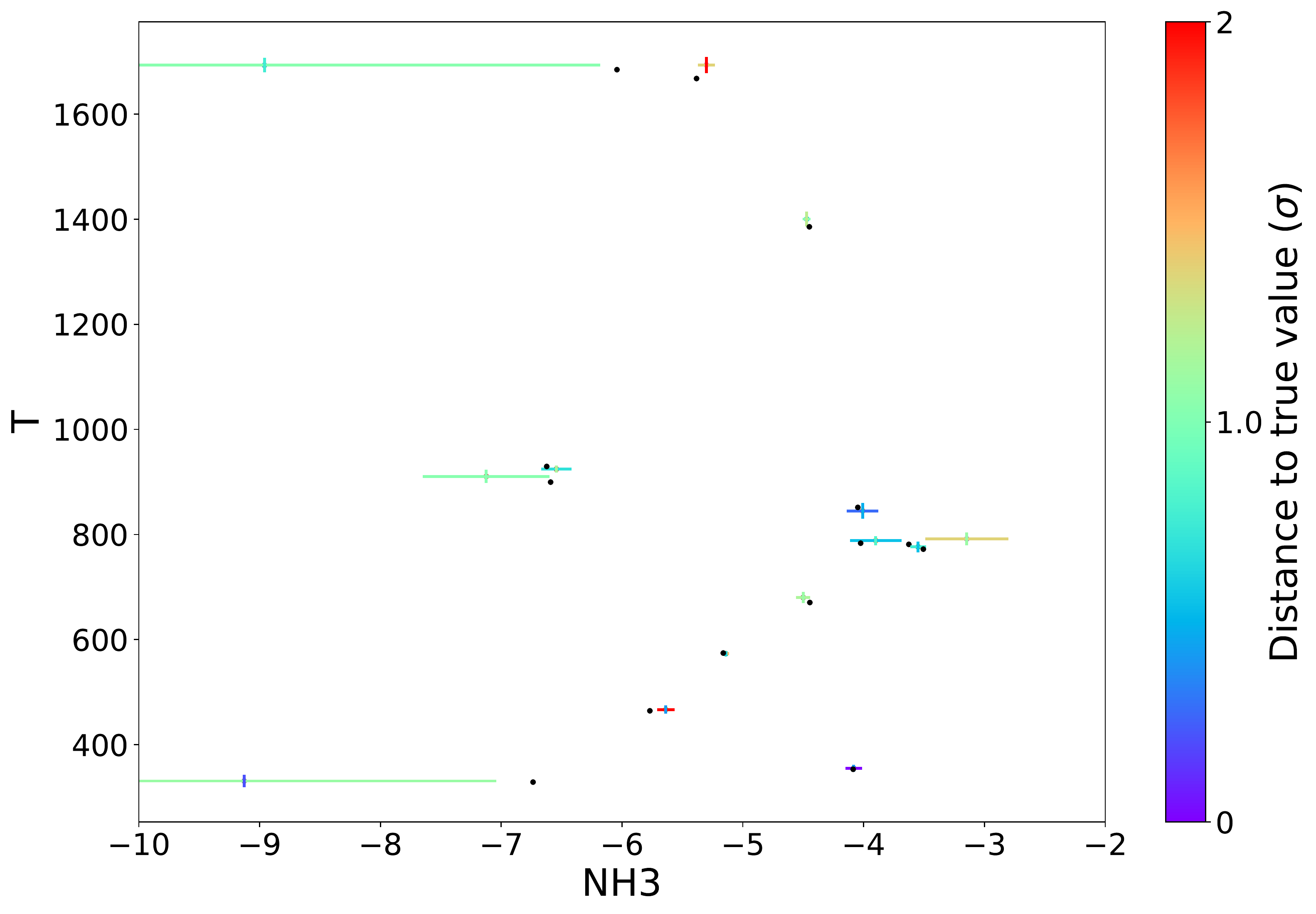}}
\caption{Correlations maps obtained for Ariel Tier-3 scattered spectra: NH$_3$-$T$. }
\label{fig:tier3_nh3}
\end{figure*}

For completeness, we performed additional retrievals for 14 benchmark planets in Tier-3 mode \citep{edwards_ariel}. The benchmark planets achieve a high signal-to-noise ratio in a very limited number of transits and are re-observed at different times to allow for temporal and spatial variability studies. In the  examples presented here, we combined five transit observations to reach the required signal-to-noise for Tier-3 \citep{edwards_ariel, Tinetti_ariel}. The retrievals were performed on the scattered spectra and are illustrated in Figure \ref{fig:spectra_tier3}. The retrieval maps for the 14 Ariel Tier-3 cases are reported in Appendix (Figures \ref{fig:tier3_h2o_ch4}, \ref{fig:tier3_co_co2}, \ref{fig:tier3_nh3}) and the molecular detection limits in Table \ref{tab:detection_limit}. The detection limit for Ariel Tier-3 spectra is very low, typically mixing ratios equal or smaller than $10^{-7}$ can be retrieved. Even CO   at mixing ratios of $\sim 10^{-6}$ appears to be detectable. 
Due to the limited number of studied cases, the Tier-3 detection limits reported here should be taken with caution and will be refined in a separate paper dedicated to the study of Tier-3 planets.

\begin{figure*}
\centering
\centerline{\includegraphics[width=0.83\textwidth]{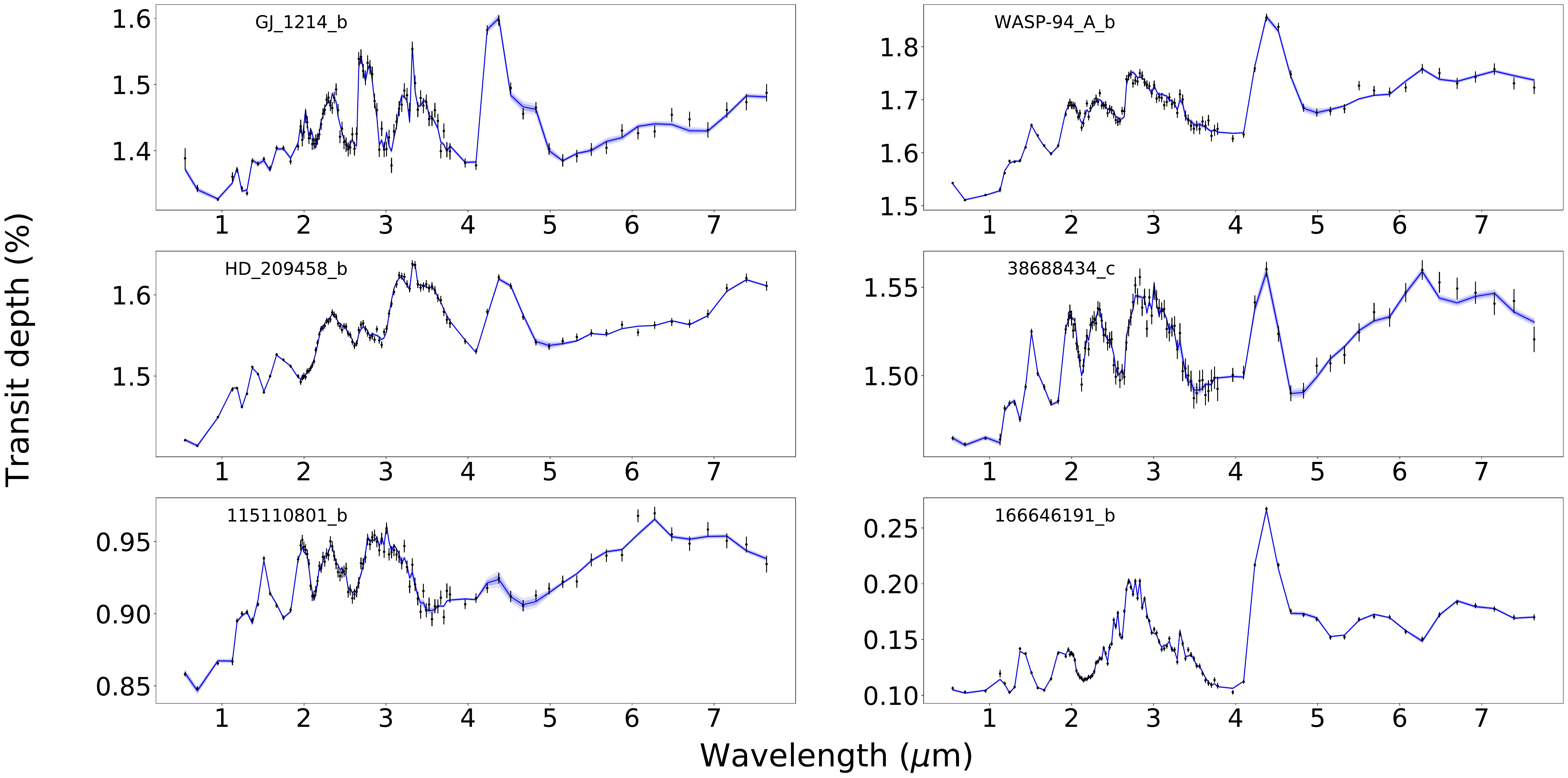}}
\caption{Examples of Ariel Tier-3 spectra and fits for benchmark  planets. The simulated observed spectra are scattered around their true value.}
\label{fig:spectra_tier3}
\end{figure*}

\subsection{Biased sample: linear water-temperature trend}

\begin{figure*}
\centering
\includegraphics[width=0.70\textwidth]{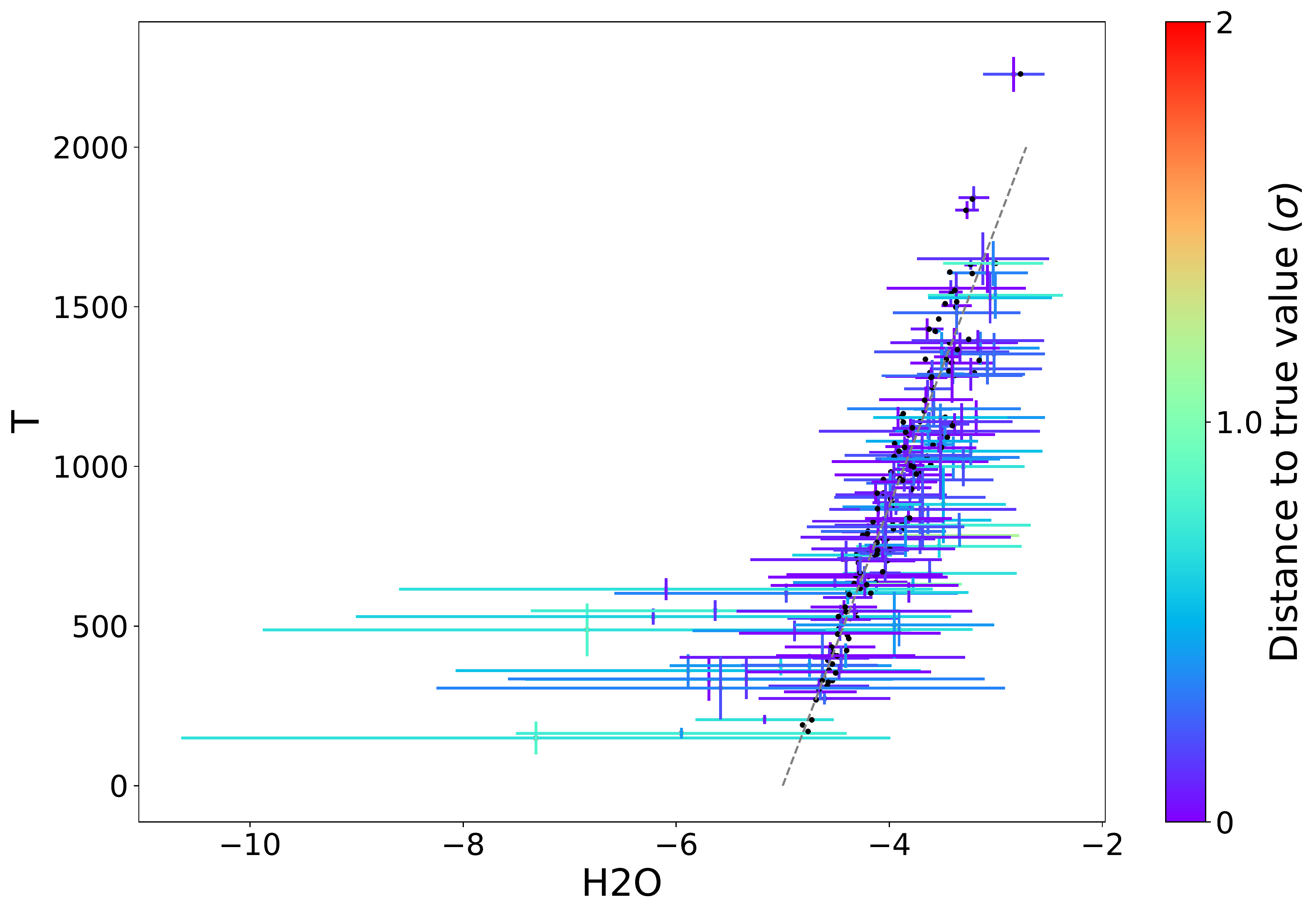}
\includegraphics[width=0.70\textwidth]{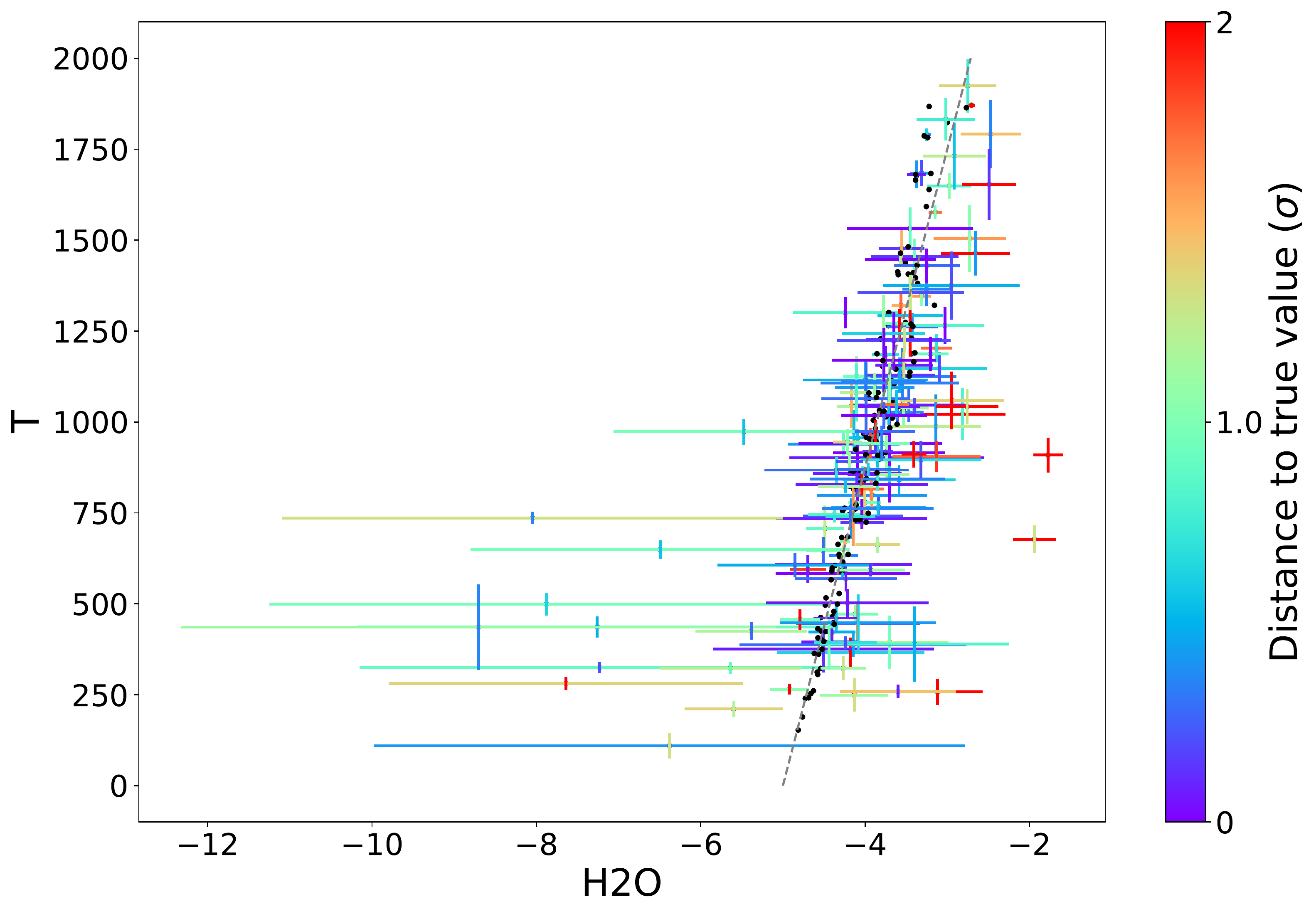}
\caption{Biased sample: linear water-temperature trend.  Top: retrieved water-temperature map from the non-scattered spectra. Bottom: retrieved water-temperature map from the scattered spectra. We show the retrieved 1-$\sigma$ error bars on the retrieved parameters. The colour-scale represents the distance to the true value in units of  1-$\sigma$. The dashed grey line indicates the input trend.}
\label{fig:params_biased}
\end{figure*}

When we imposed an arbitrary linear trend between the water abundance and the effective temperature, we obtained the water-temperature map  shown in Figure \ref{fig:params_biased}. Here the imposed trend is easily recovered by our retrieval analysis. Both scattered and unscattered spectra allow to recover the imposed trend down to water abundances $\sim 10^{-6}$. In the scattered example, a few cases have larger departures from the true value compared to the non-scattered one but this does not affect the conclusions on the entire population. Additionally, we note that this analysis has been done without retrieval fine tuning.

\subsection{Biased sample: equilibrium chemistry atmospheres}\label{sec:eq_section}

\begin{figure*}
\centering
\includegraphics[width=0.70\textwidth]{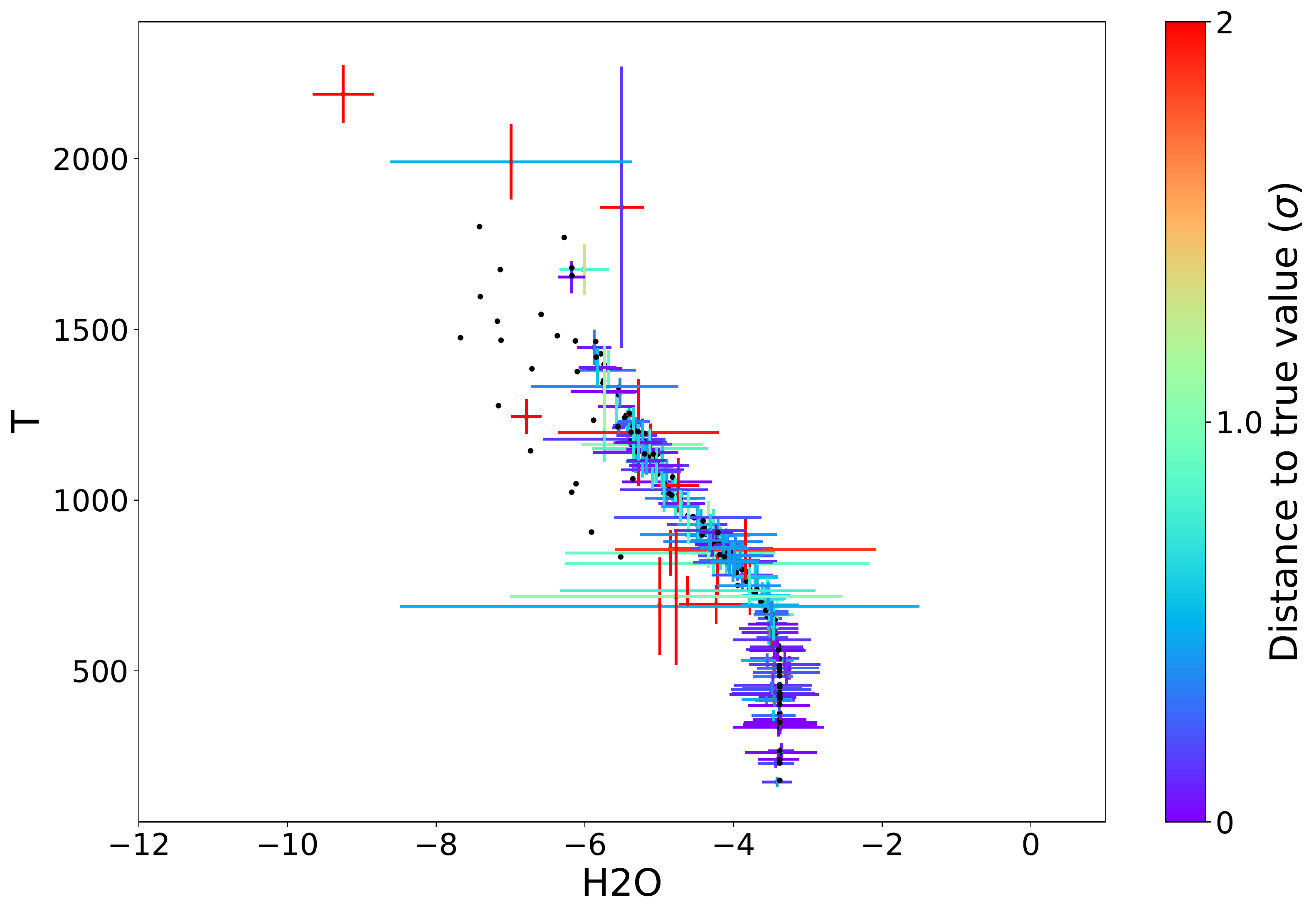}
\includegraphics[width=0.70\textwidth]{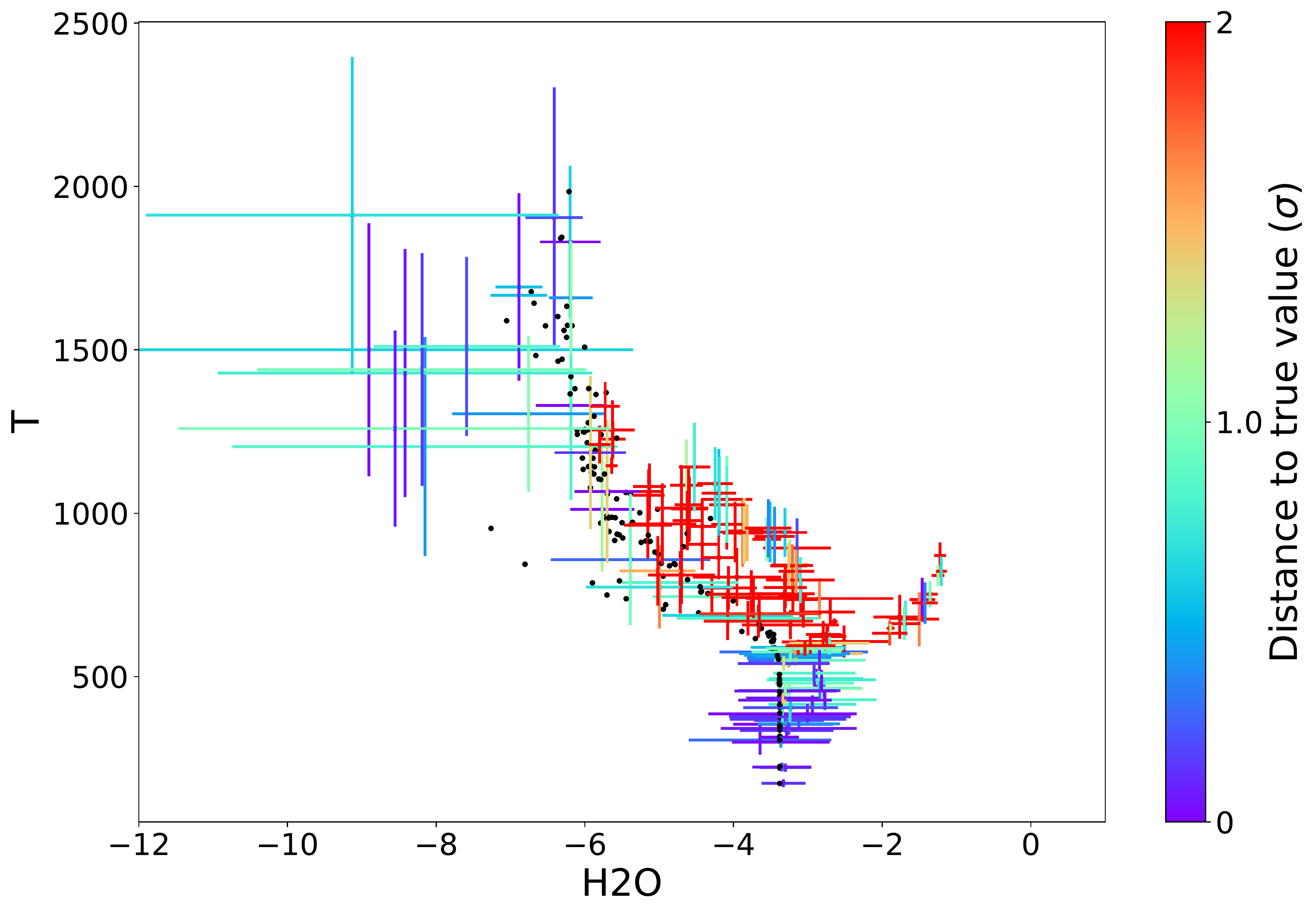}
\caption{Biased sample:  equilibrium chemistry atmospheres.  Correlation map of the retrieved abundance of water and the temperature. Results obtained with equilibrium chemistry retrievals (top) and with free, constant  with altitude chemistry retrievals (bottom). We show the retrieved 1-$\sigma$ error bars on the retrieved parameters. The colour-scale represents the distance to the true value in units of  1-$\sigma$. }
\label{fig:params_ace}
\end{figure*}

When an equilibrium chemistry model was used for both the forward model and the retrievals, we obtained the water-temperature map shown in Figure \ref{fig:params_ace}, top, where the trend is very accurately and precisely recovered. Since the molecular abundances are varying with altitude, the values stated correspond to the average weighted by the atmospheric contribution function (the optical depth variations collapsed over wavelengths).   Being  the model generating and retrieving the data  the same, this is an optimistic result, as we should not expect all atmospheres to satisfy the equilibrium chemistry assumption.

\begin{figure*}[h]
\centering
\includegraphics[width=0.81\textwidth]{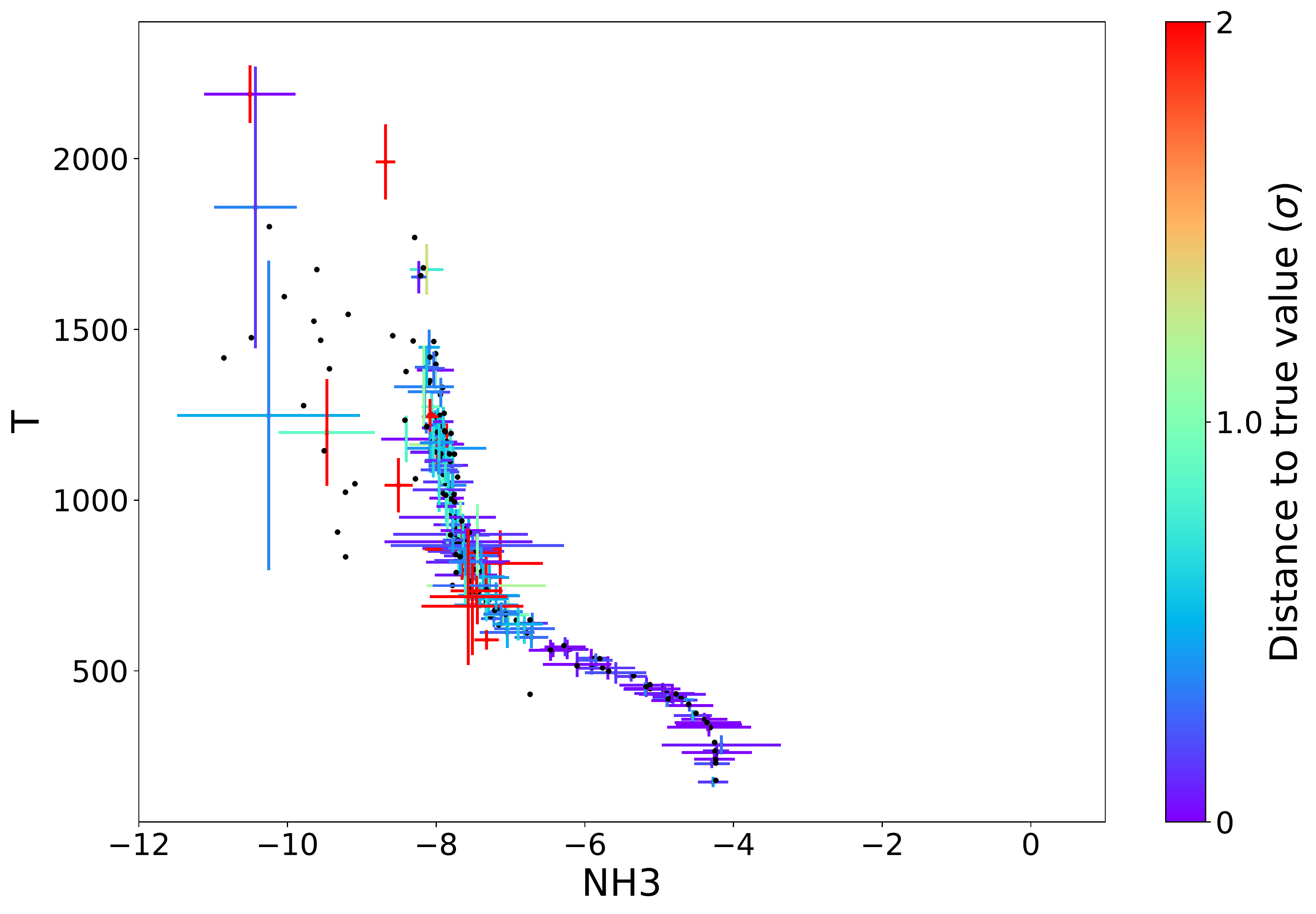}
\includegraphics[width=0.81\textwidth]{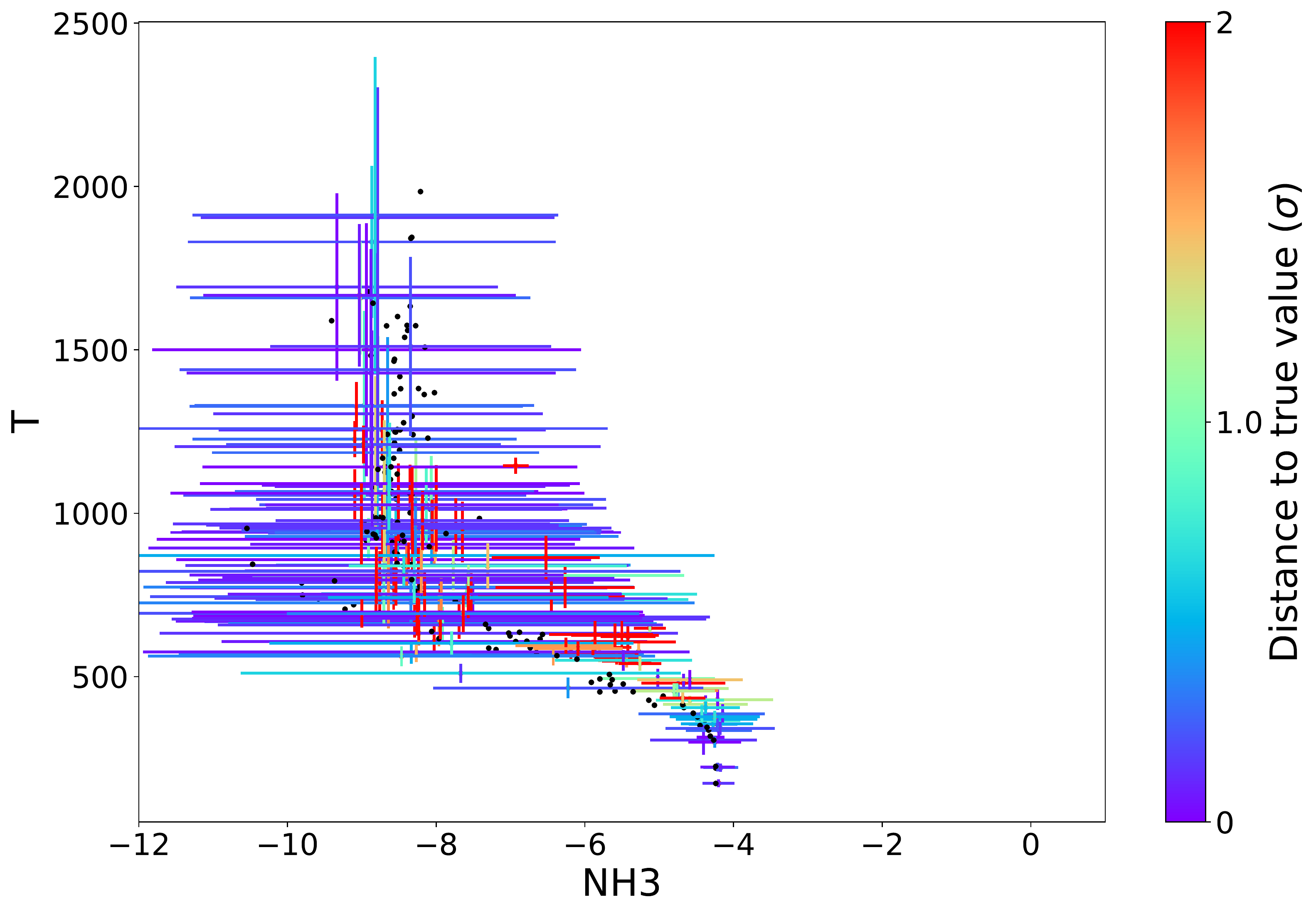}
\caption{Biased sample:  equilibrium chemistry atmospheres.  Correlation map of the retrieved abundance of NH$_3$ and the temperature, with the 1-$\sigma$ retrieved error bars. Results obtained with equilibrium chemistry retrievals (top) and with free, constant with pressure chemistry retrievals (bottom). The colour-scale represents the distance to the true value in units of  1-$\sigma$.
}
\label{fig:aceNH3}
\end{figure*}

Also the free, constant  with pressure chemistry retrievals (Figure \ref{fig:params_ace}, bottom) allow to recover the equilibrium chemistry trend. The retrieved parameters, however, have large distances from the true value, in some cases the offsets are greater than 2$\sigma$, meaning that the model confidently recovers a biased value. This behaviour, also present in  other chemical species (see Appendix Figures \ref{fig:aceCH4}, \ref{fig:aceCO}, \ref{fig:aceCO2}, \ref{fig:aceNH3}), is particularly noticeable for temperatures between 600K and 1100K: this region is known to exhibit large chemical gradients with altitude as the balance in the CH$_4$/CO reaction changes. These variations in the chemical profiles  cannot be captured by our  simplistic constant chemistry retrieval model. 

It has been shown in \cite{Changeat_2019_2layer} that Ariel and JWST will be sensitive to chemical vertical gradients and that retrieval techniques such as the two-layer parametrisation would be essential for the analysis of these next generation spectra. 

\begin{figure*}
\centering
\includegraphics[width=0.5\textwidth]{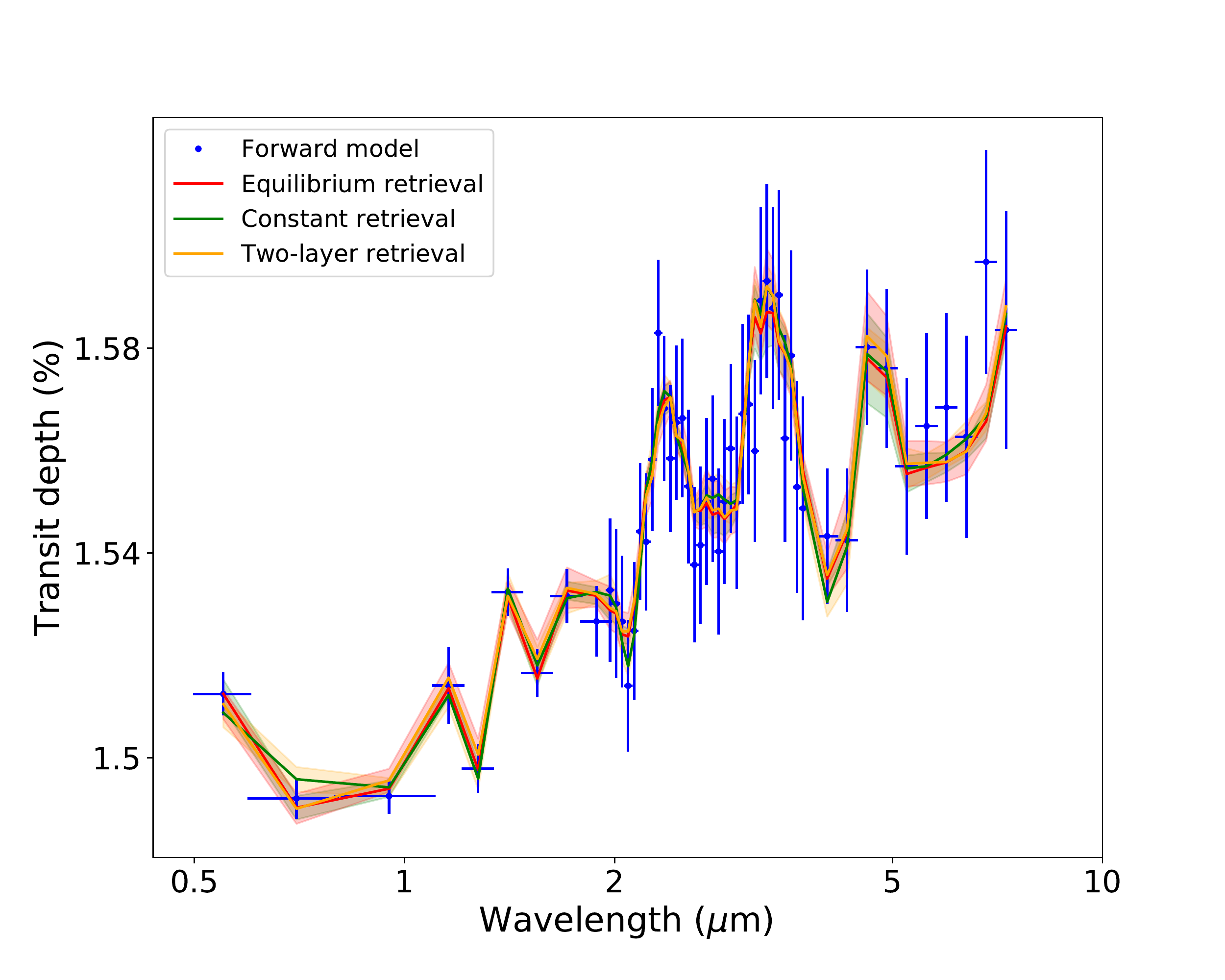}
\includegraphics[width=0.48\textwidth]{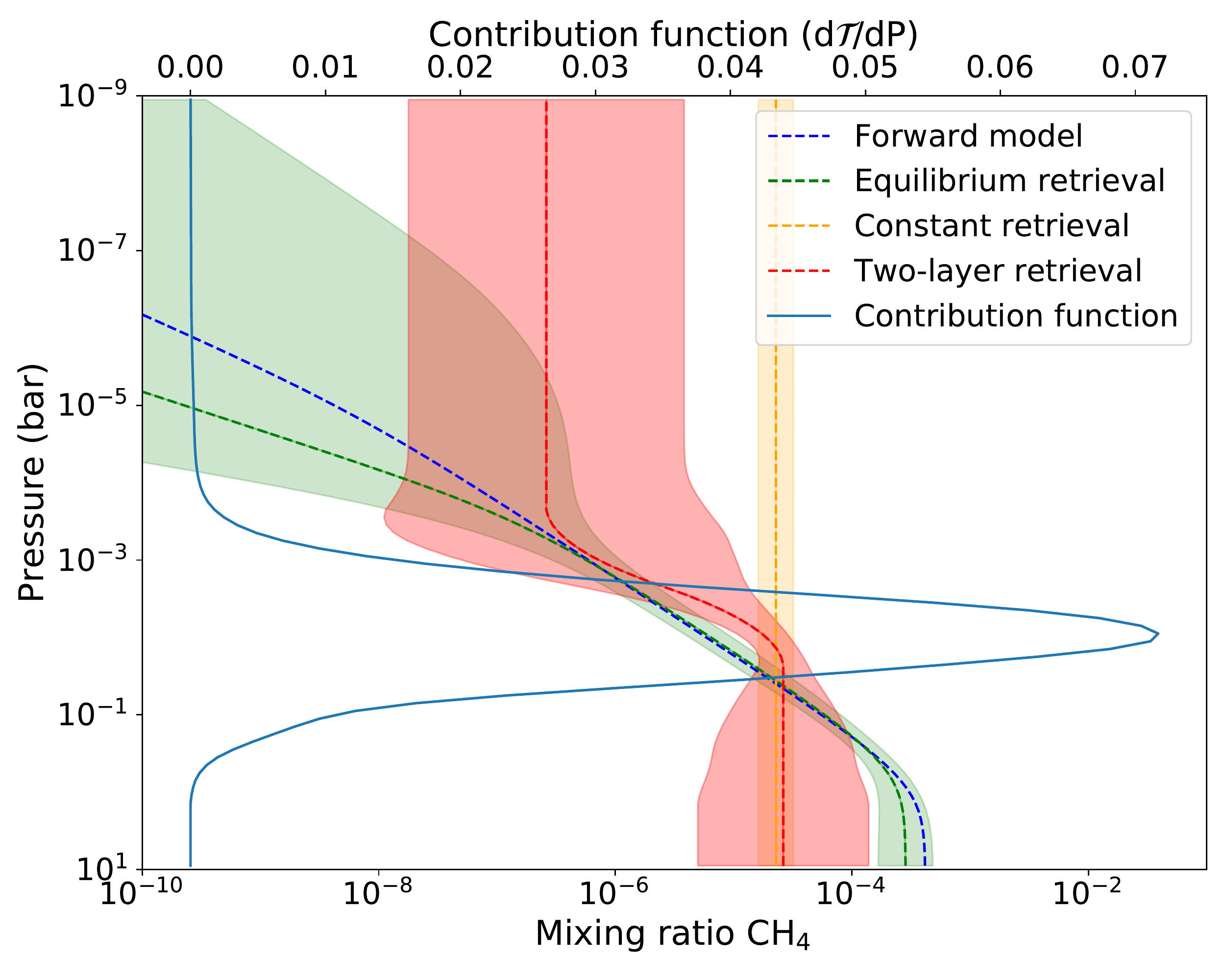}
\caption{Results for our retrievals with three different chemical profiles (equilibriun, constant with altitude and two-layer). The input forward model is taken from the Alfnoor run with equilibrium chemistry. Left: Simulated observations and retrieval best fit models; Right: Comparison of the retrieved CH$_4$ profiles. The contribution function in the atmosphere , corresponding to d$\mathcal{T}$/dP, is also provided. The global log evidence, which qualifies the preference shown by the data for a given model, is 400 for the equilibrium model , 397.5 for the two-layer retrieval and only 395 for the constant chemistry retrieval. In comparing models, a difference of 2 indicate a strong preference towards the model of higher value \citep{Kass1995bayes}. We calculate the abundances weighted by the contribution function (log) to be:$-5.29$ for the forward model; $-5.27\pm0.21$ for the equilibrium model; $-4.64 \pm 0.15$ for the constant chemistry model; $-4.96 \pm 0.39$ for the two-layer model.}
\label{fig:params_compar}
\end{figure*}

We show in Figure (\ref{fig:params_compar}) a comparison between the various retrieval techniques:  the two-layer parametrisation \citep{Changeat_2019_2layer} well captures the departure of the methane  profile from the constant with altitude case without strong prior assumptions, as opposed to the case of the equilibrium chemistry retrieval.

\section{Discussion} \label{sec:discussion}

 In all simulated cases, retrieval analyses were performed without any fine tuning. Also our simulations are  simplified compared to real atmospheres, which are expected to  have disequilibrium effects, 3D effects and other complexities. 

Recently, self-consistent methods, such as the equilibrium chemistry retrieval adopted in a few examples here, have been implemented in retrieval tools. Embedding these chemical schemes in atmospheric retrievals is very tempting as they allow to describe complex chemistry while maintaining a low dimensionality.  However, we should be careful in using these techniques to interpret unknown atmospheres,  as they do not reflect the information content of the observed spectra. In other terms, if the assumptions made by the retrieval model are not correct, the results will likely be biased  \citep{Miller_GJ1214_clouds, Rocchetto_biais_JWST, Agundez_2012_eqchem, Changeat_2019_2layer}. This issue has been discussed in the literature and should always be remembered when using such techniques. 

Other approaches which let the chemical species assume arbitrary values, may allow to discover unexpected trends in the data. However, the model complexity should be adapted to the data, which is not known a priori. A too simplistic model will tend to be biased, while a too complex model will tend to overfit. In this paper, Section \ref{sec:eq_section} highlighted a case where the free constant with pressure chemistry retrieval did not adequately describe the input chemical profiles (which were using equilibrium chemistry), thus biasing our results. A more sophisticated  description of the chemical profiles in retrievals is presented in \cite{Changeat_2019_2layer}.

 We illustrate this point by comparing different chemical schemes on an observed spectrum taken from our previously made equilibrium chemistry dataset. Figure \ref{fig:params_compar} demonstrates that all three chemical schemes (equilibrium, constant, two-layer) are able to match the observed spectrum. The contribution function (solid blue line on the right figure) shows how the models try to reproduce the input abundances for $CH_4$ in the region where the contribution function is maximum. The equilibrium and two-layer scenarios are better describing the input profiles in general, while the retrieved uncertainties are more representative. The retrieved constant chemical profile with altitude only averages the input CH$_4$ abundance, providing limited details on the atmospheric chemical processes. As expected, we find that the input retrieved weighted abundance is best approximated by the equilibrium model, since this is the same model used to generate the observation (values are stated in Figure \ref{fig:params_compar}). The constant with pressure chemistry model is overconfident and is more than 3$\sigma$ offset to the true value. For the two-layer, the true abundance is within the error bars of the retrieved value. The behaviour seen in this example explains the large distances to the true value and the general overconfidence in the retrieved chemistry of our free constant with altitude scenario in Figure \ref{fig:params_ace}.

\section{Conclusion} \label{sec:conclusion}

This work assessed the capabilities of Ariel to identify chemical trends -- if present -- in exoplanet populations through the study of their atmospheres. 
We developed a dedicated software, Alfnoor,  to perform atmospheric retrievals on the entire Ariel list of planetary candidates. Among the key results obtained, we found the detection limits  for H$_2$O, CH$_4$,  CO$_2$ and NH$_3$ to be $\sim 10^{-6}$ in the case of Tier-2 and $< 10^{-7}$ in the case of Tier-3 transit observations. CO, though, has higher detection thresholds,  i.e. $\sim 10^{-4}$ for Tier-2 observations and  $\sim 10^{-6}$ for Tier-3.

We also confirmed the potentials of Ariel to recover chemical trends in exoplanetary atmospheres. We tested correlations between chemical species and temperature and a planet population whose chemical composition is entirely determined by equilibrium chemistry. 

Limitations in our assumptions for the chemistry, temperature and cloud models imply that additional work  still needs to be done to fully understand the degeneracies associated with these techniques and  how to fully automate retrieval strategies. In the future, we aim to simulate more realistic scenarios using  self consistent forward models (e.g. including disequilibrium chemistry) and more complex thermal and cloud assumptions.  While this work was inspired by the Ariel mission, similar large scale simulations  could also help prioritising the use of other observatories from space and the ground and provide a great tool for the preparation of observational campaigns.

\section{Acknowledgement}

This project has received funding from the European Research Council (ERC) under the European Union's Horizon 2020 research and innovation programme (grant agreement No 758892, ExoAI) and under the European Union's Seventh Framework Programme (FP7/2007-2013)/ ERC grant agreement numbers 617119 (ExoLights). Furthermore, we acknowledge funding by the Science and Technology Funding Council (STFC) grants: ST/K502406/1, ST/P000282/1, ST/P002153/1, ST/T001836/1 and ST/S002634/1. ASI grant n. 2018.22.HH.O. UCL London-Rome Cities Partnerships Program.

\newpage
\bibliographystyle{aasjournal}
\bibliography{main}

\section{Appendix}

\begin{figure*}[h]
\centerline{\includegraphics[width=0.81\textwidth]{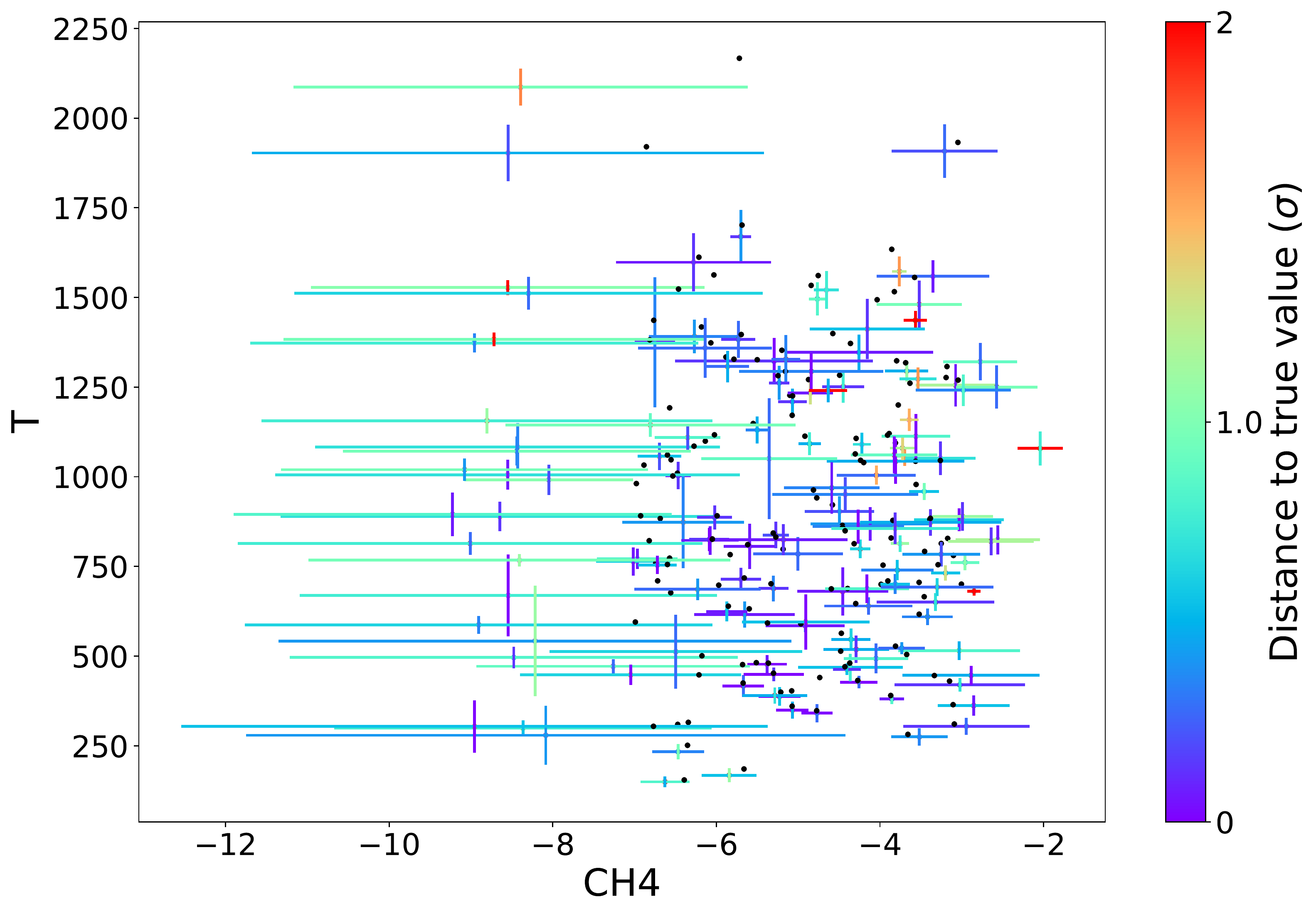}}
\centerline{\includegraphics[width=0.81\textwidth]{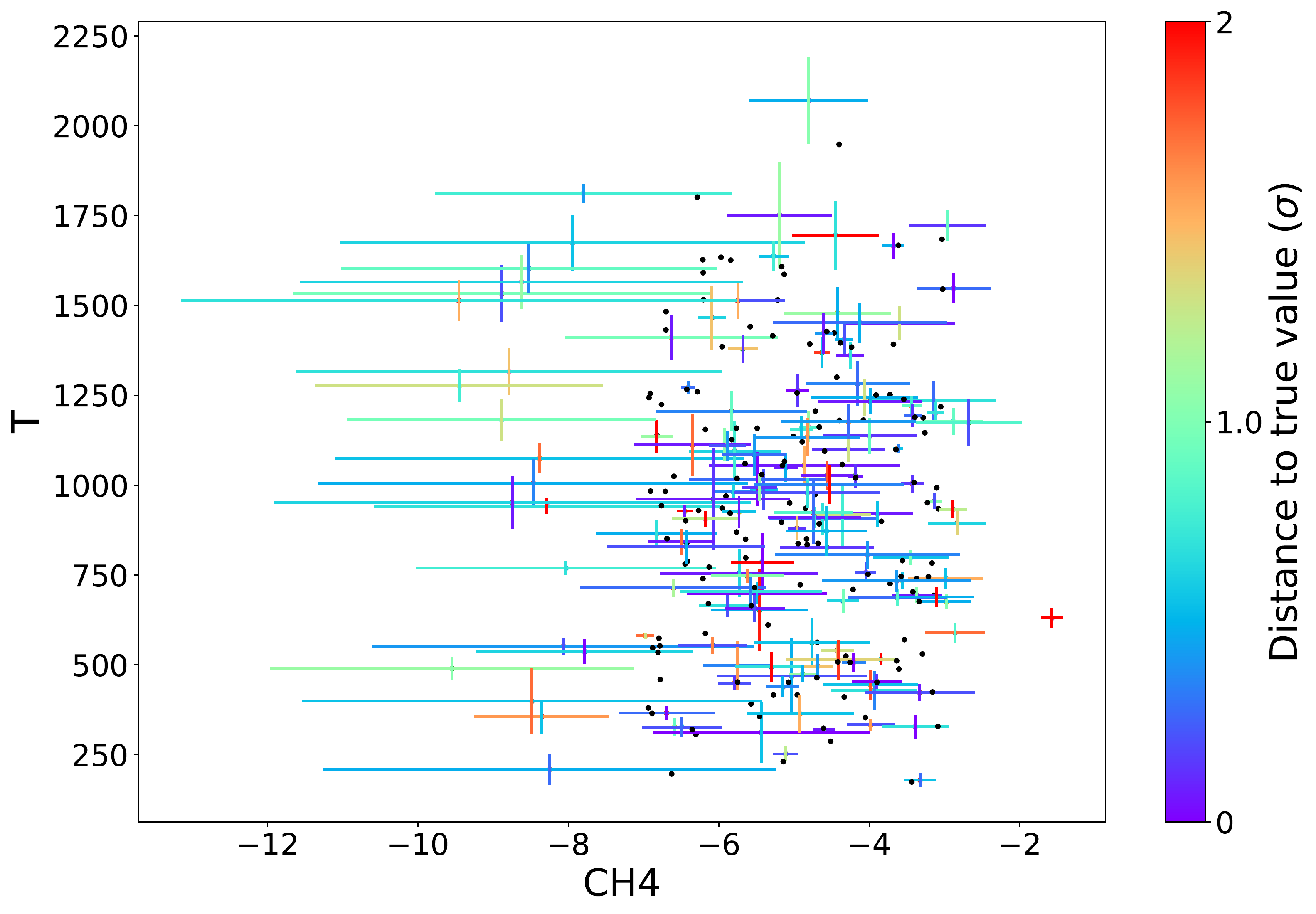}}
\caption{Unbiased sample: correlation map between the temperature and the retrieved abundances of CH$_4$, with the 1-$\sigma$ retrieved error bars. The colour-scale represents the distance to the true value in units of  1-$\sigma$. Top: Non-scattered spectra. Bottom: Scattered spectra.
}
\label{fig:sc_ch4}
\end{figure*}

\begin{figure*}[h]
\centerline{\includegraphics[width=0.81\textwidth]{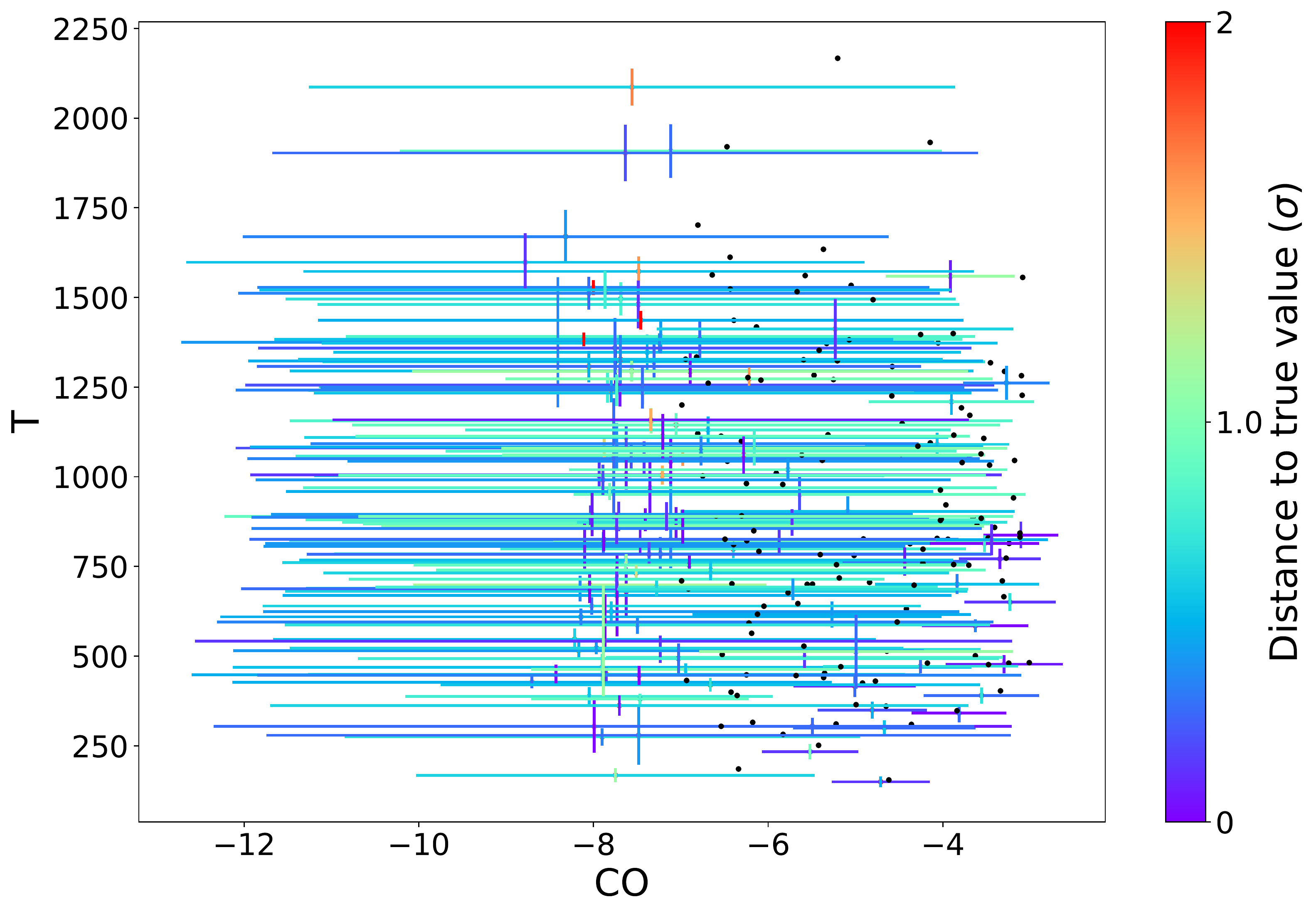}}
\centerline{\includegraphics[width=0.81\textwidth]{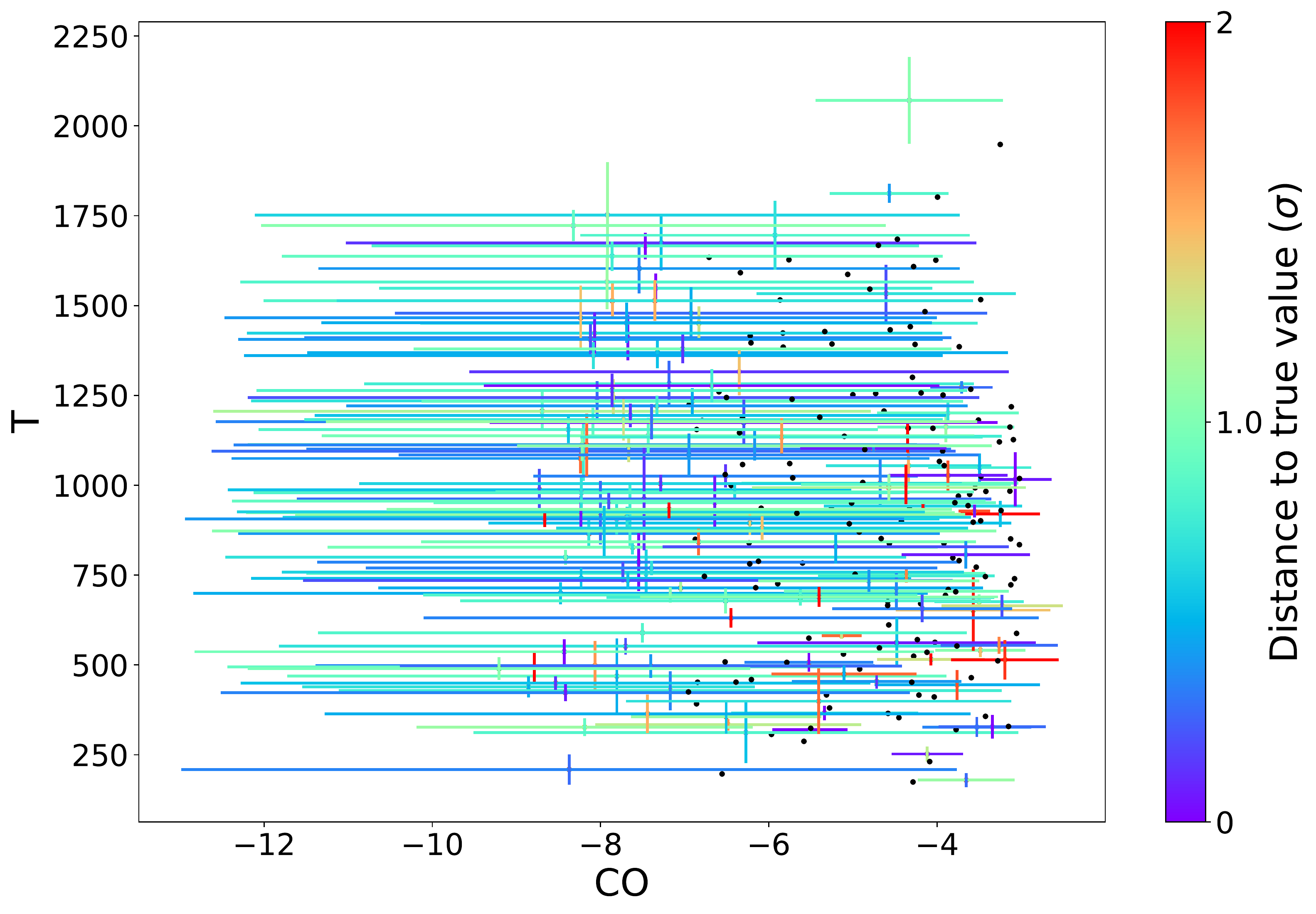}}
\caption{Unbiased sample: correlation map between the temperature and the retrieved abundances of CO, with the 1-$\sigma$ retrieved error bars. The colour-scale represents the distance to the true value in units of  1-$\sigma$. Top: Non-scattered spectra. Bottom: Scattered spectra.}
\label{fig:sc_co}
\end{figure*}

\begin{figure*}[h]
\centerline{\includegraphics[width=0.81\textwidth]{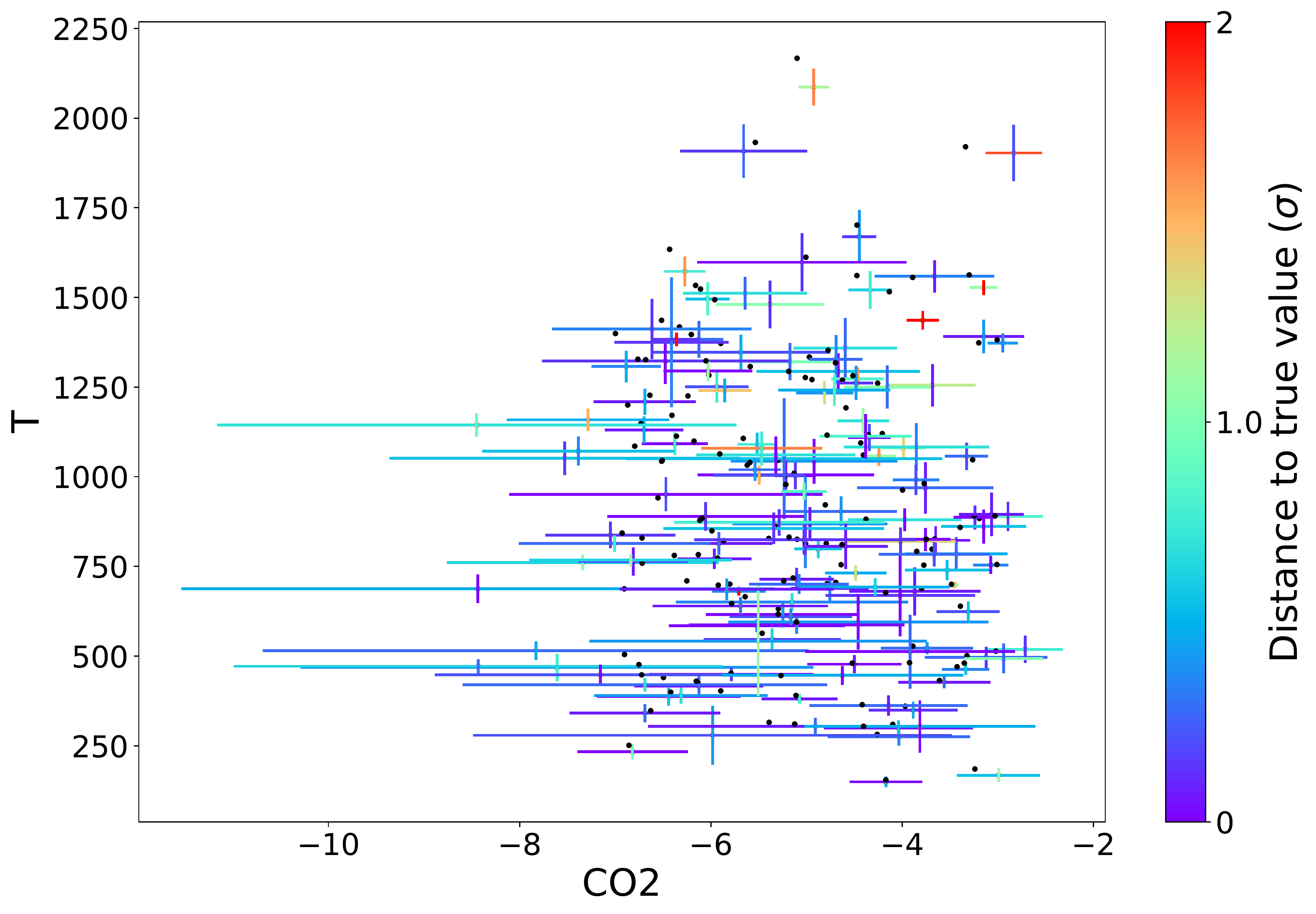}}
\centerline{\includegraphics[width=0.81\textwidth]{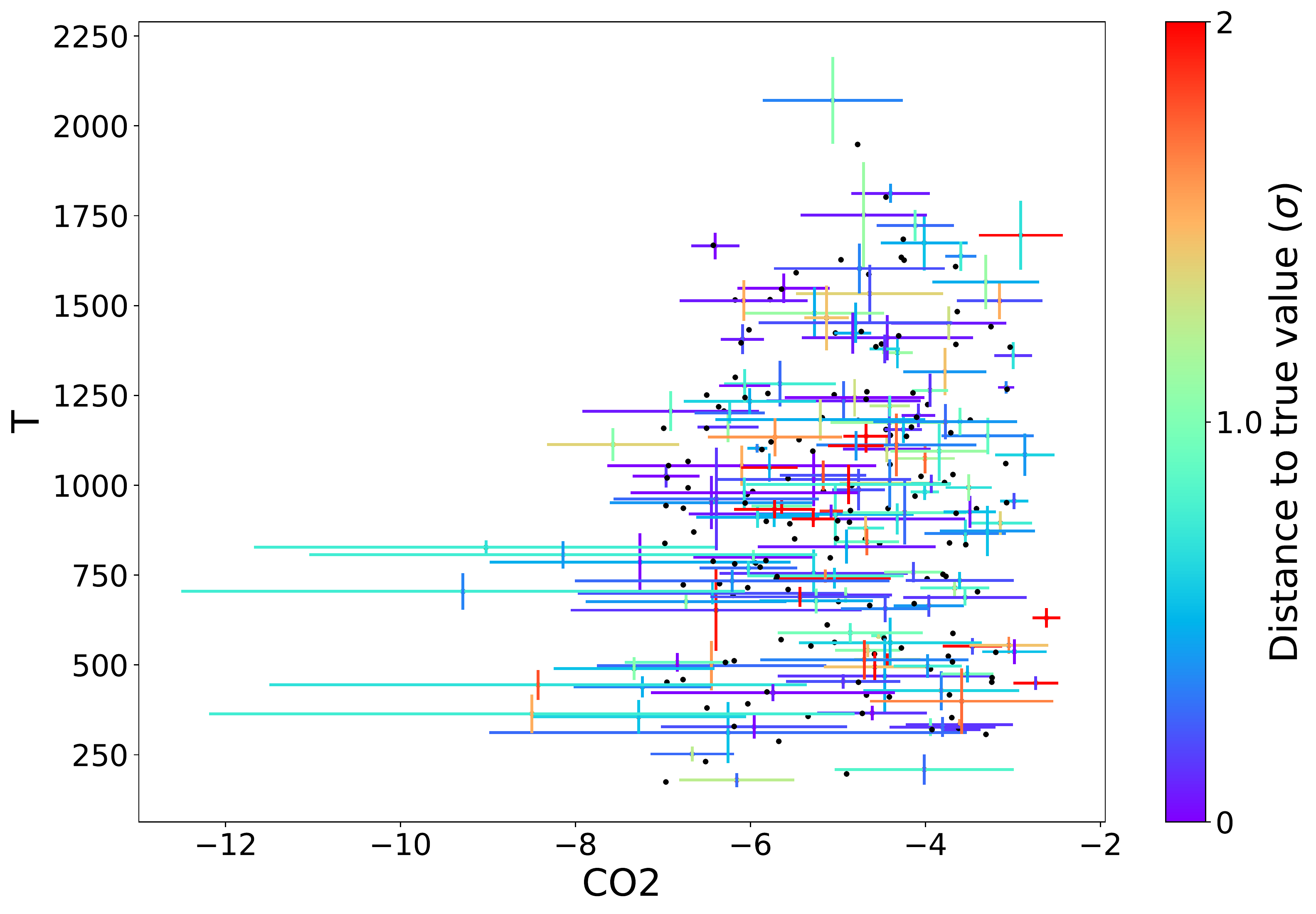}}
\caption{Unbiased sample: correlation map between the temperature and the retrieved abundances of CO$_2$, with the 1-$\sigma$ retrieved error bars. The colour-scale represents the distance to the true value in units of  1-$\sigma$. Top: Non-scattered spectra. Bottom: Scattered spectra.  }
\label{fig:sc_co2}
\end{figure*}

\begin{figure*}[h]
\centering
\centerline{\includegraphics[width=0.81\textwidth]{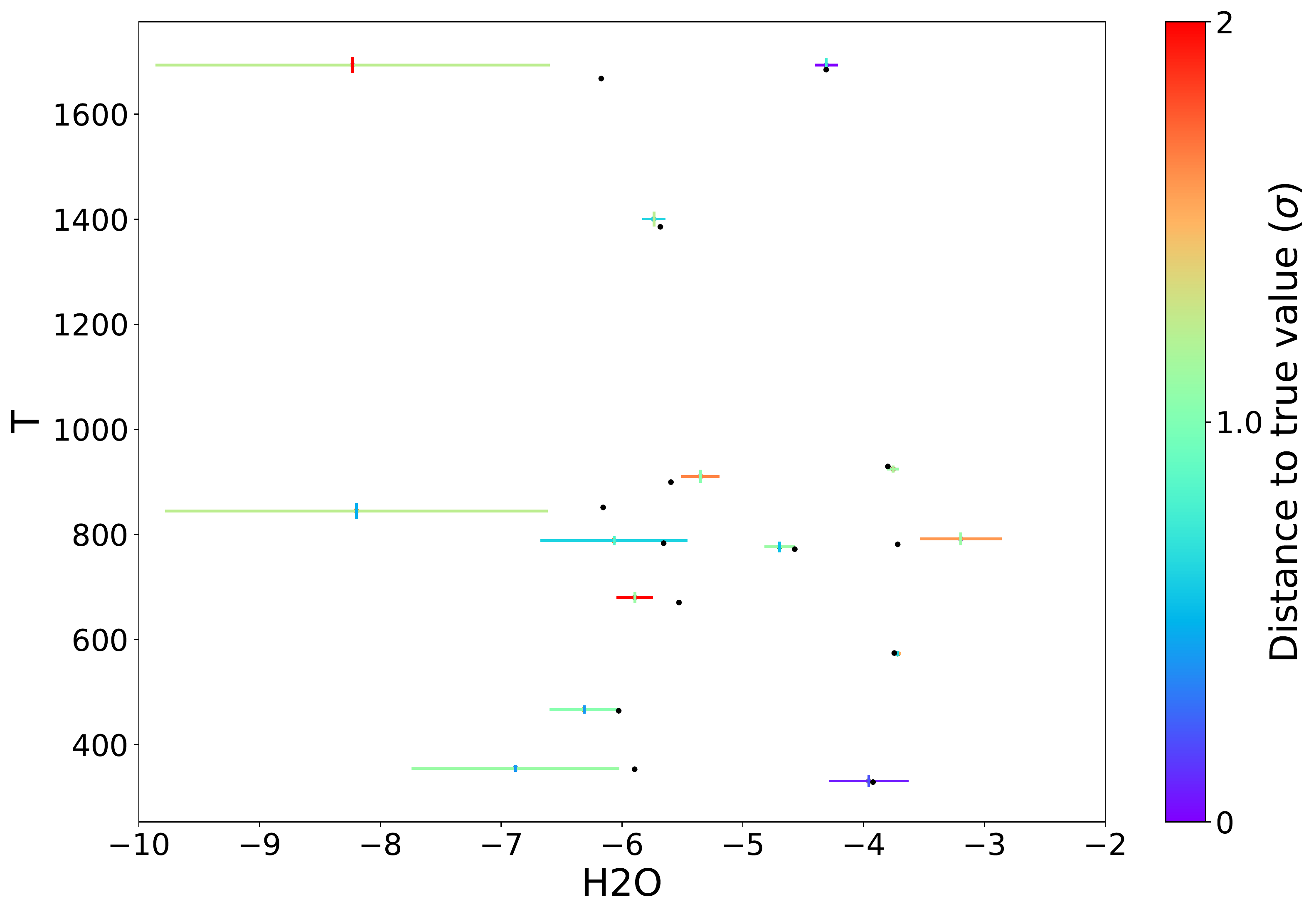}}
\centerline{\includegraphics[width=0.81\textwidth]{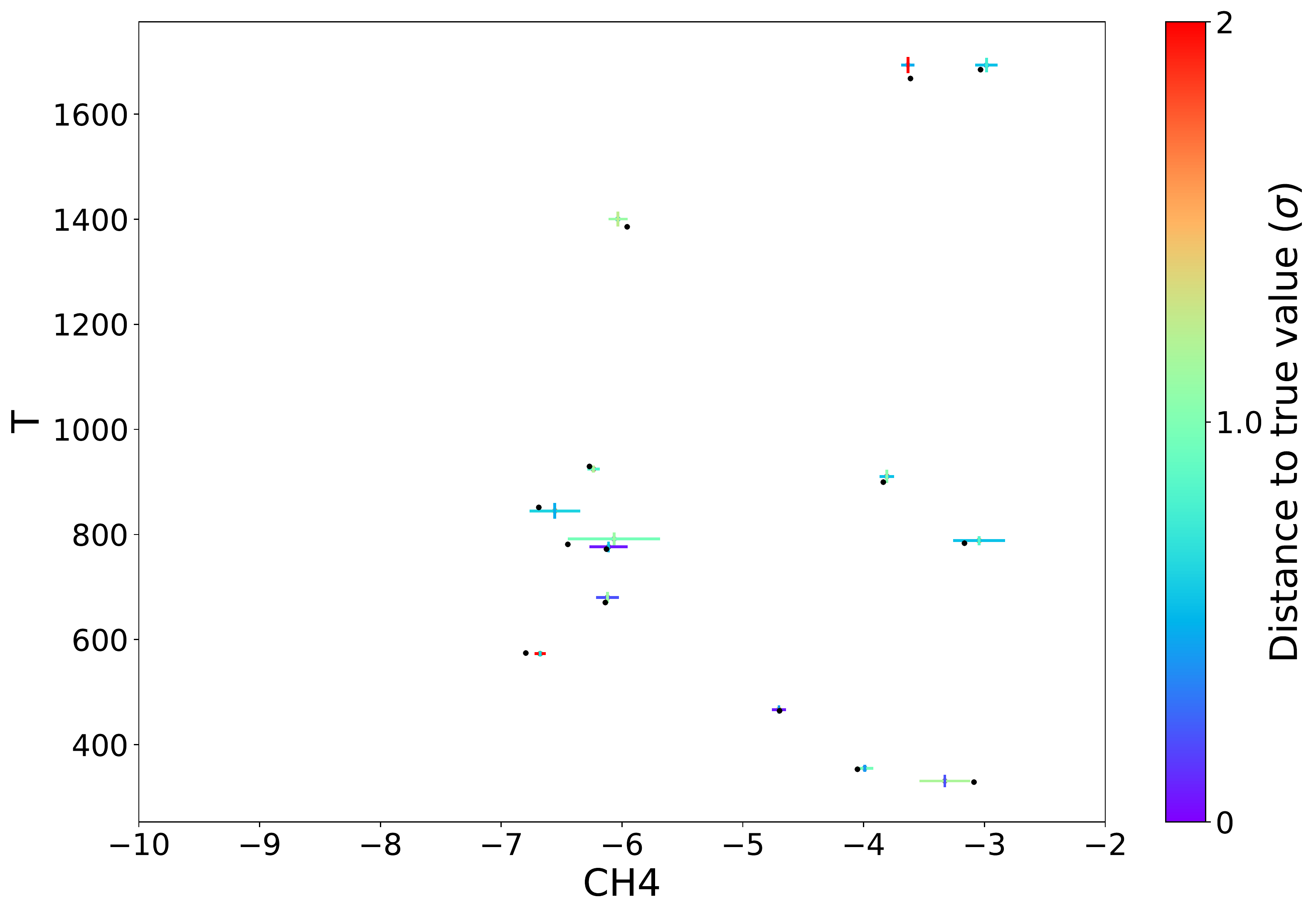}}
\caption{Correlations maps obtained for Ariel Tier-3 scattered spectra:  H$_2$O-$T$ (top) and CH$_4$-$T$ (bottom). }
\label{fig:tier3_h2o_ch4}
\end{figure*}

\begin{figure*}[h]
\centerline{\includegraphics[width=0.81\textwidth]{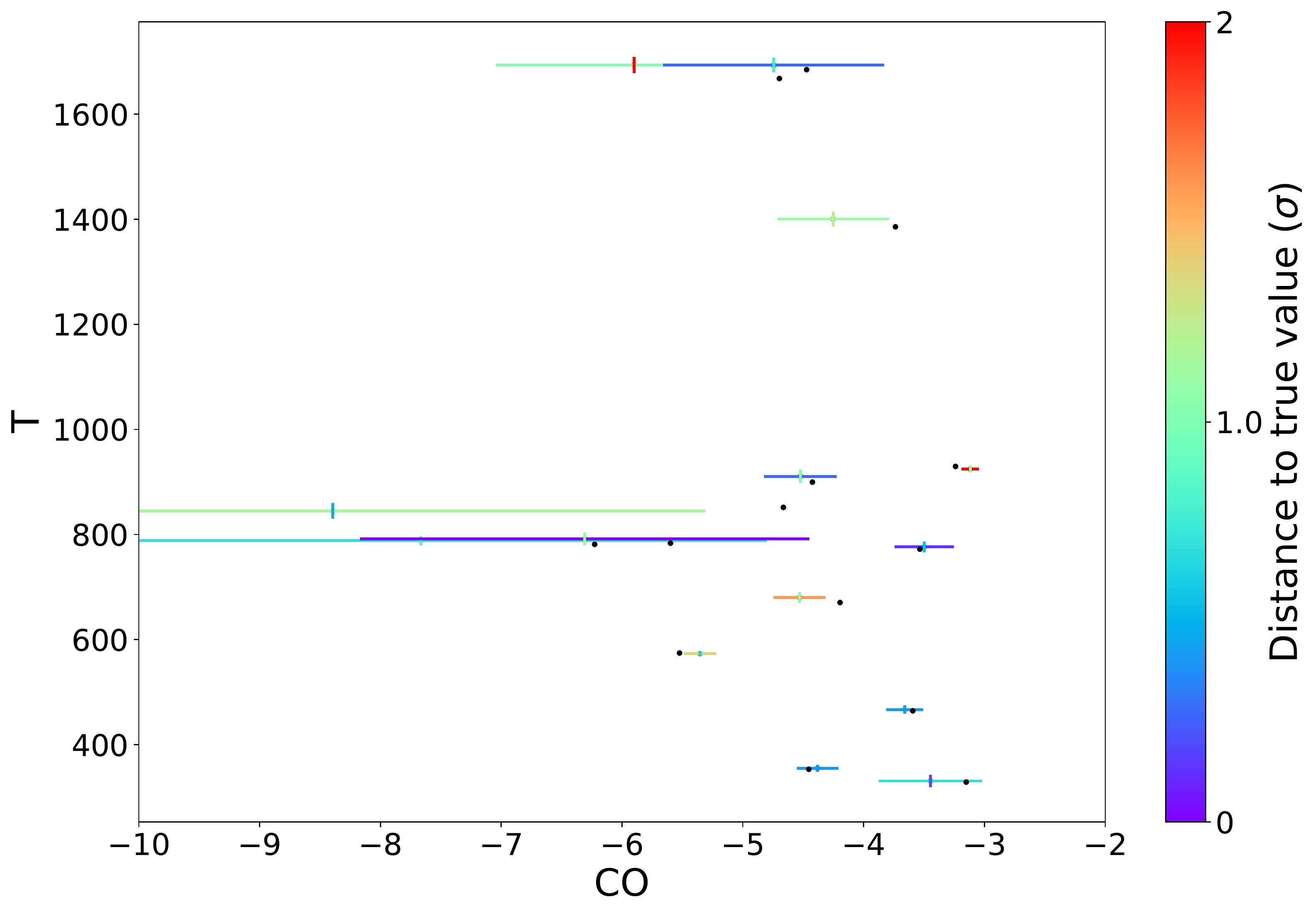}}
\centerline{\includegraphics[width=0.81\textwidth]{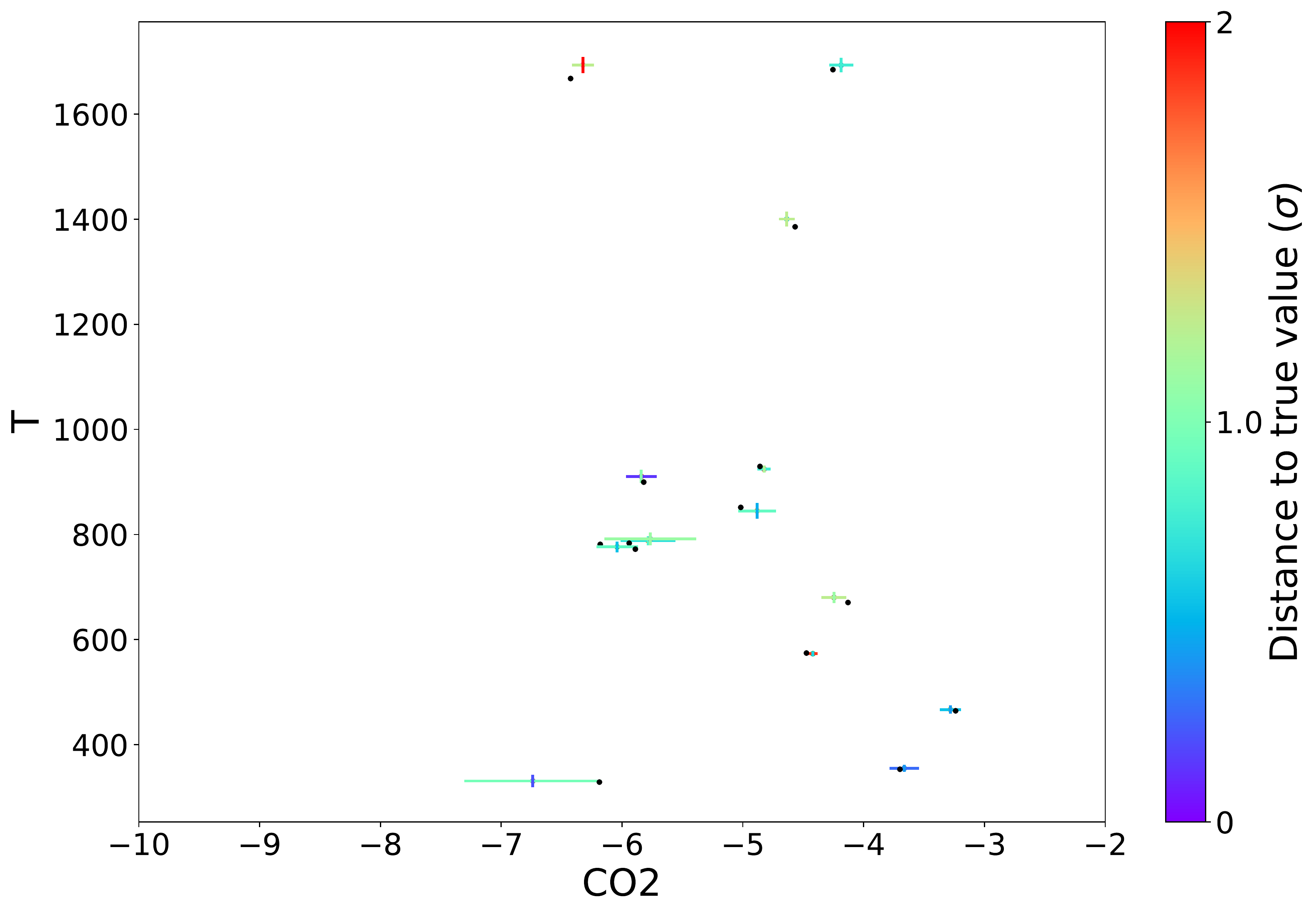}}
\caption{Correlations maps obtained for Ariel Tier-3 scattered spectra:  CO-$T$ (top) and CO$_2$-$T$ (bottom). }
\label{fig:tier3_co_co2}
\end{figure*}

\begin{figure*}[h]
\centering
\centerline{\includegraphics[width=0.81\textwidth]{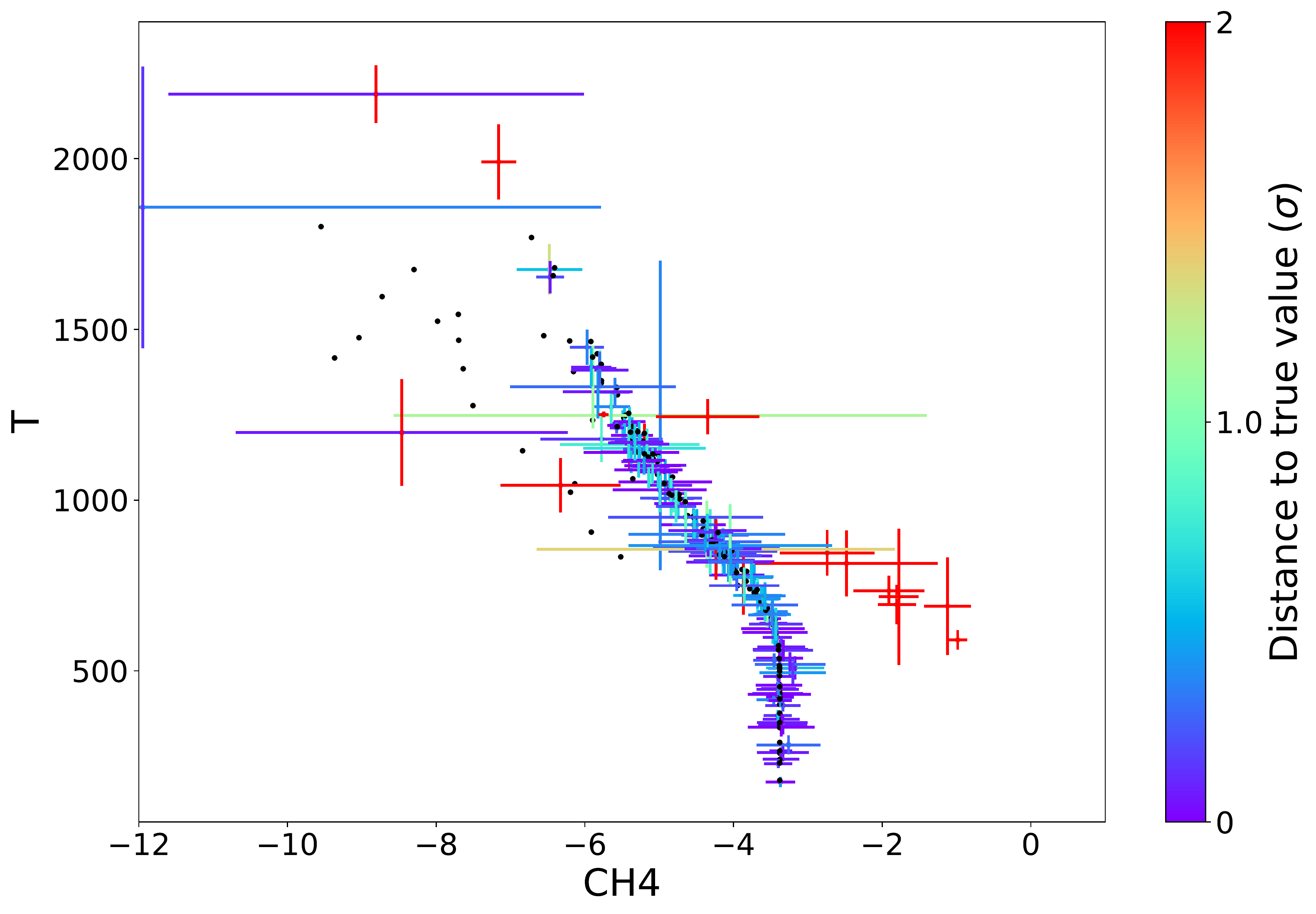}}
\centerline{\includegraphics[width=0.81\textwidth]{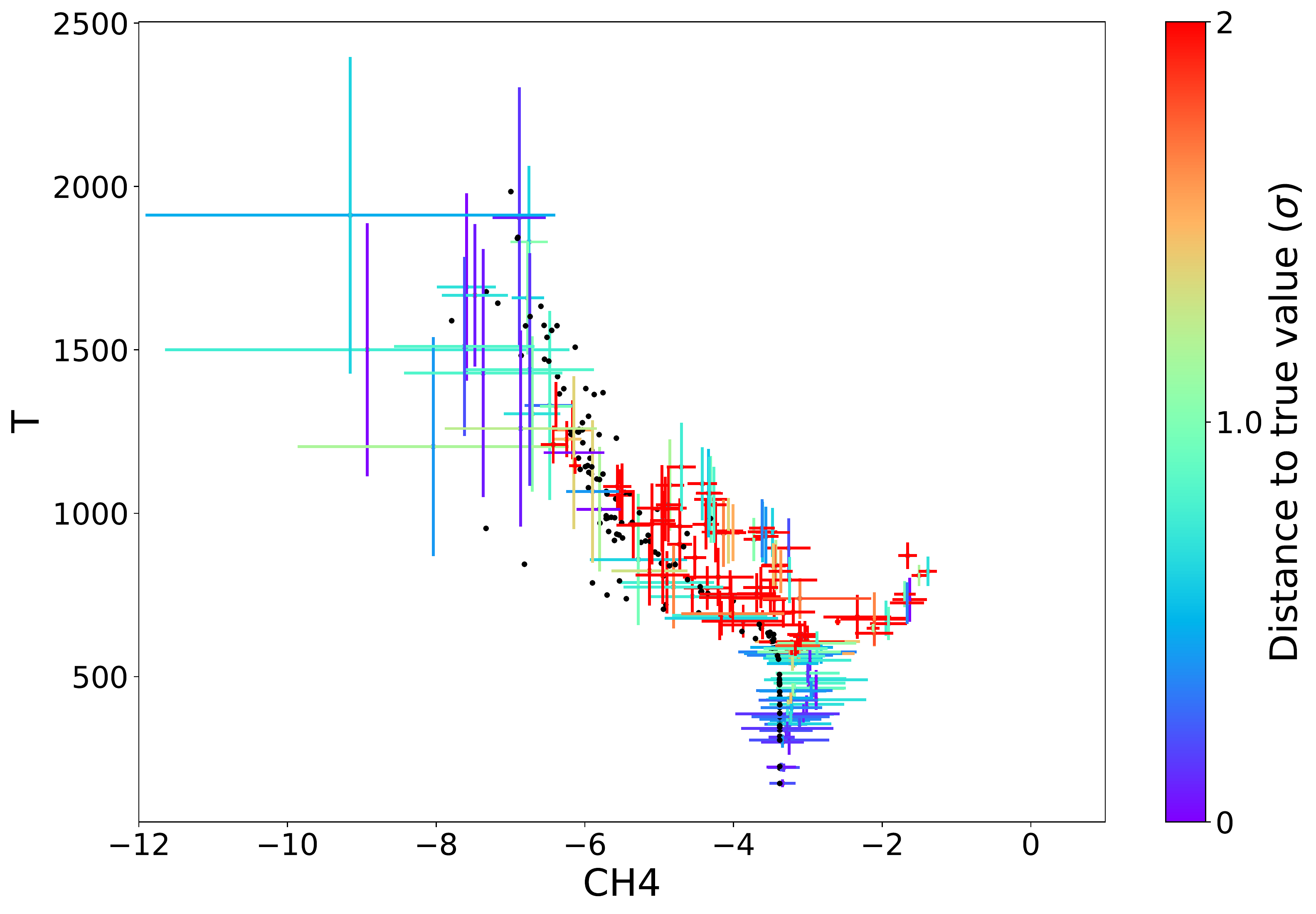}}
\caption{Biased sample:  equilibrium chemistry atmospheres.  Correlation map of the retrieved abundance of CH$_4$ and the temperature, with the 1-$\sigma$ retrieved error bars. Results obtained with equilibrium chemistry retrievals (top) and with free, constant chemistry  with pressure retrievals (bottom). The colour-scale represents the distance to the true value in units of  1-$\sigma$. }
\label{fig:aceCH4}
\end{figure*}

\begin{figure*}[h]
\centering
\centerline{\includegraphics[width=0.81\textwidth]{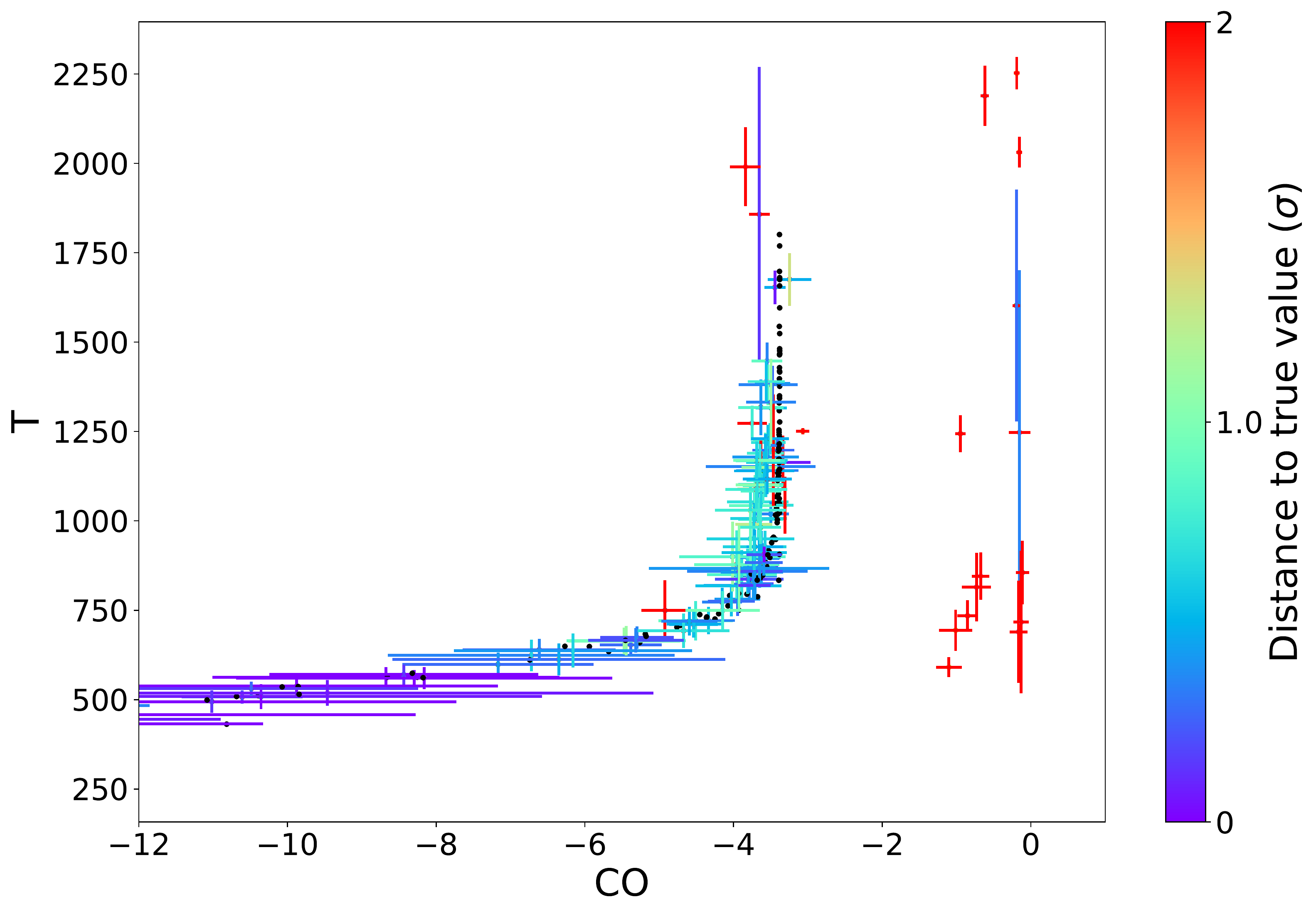}}
\centerline{\includegraphics[width=0.81\textwidth]{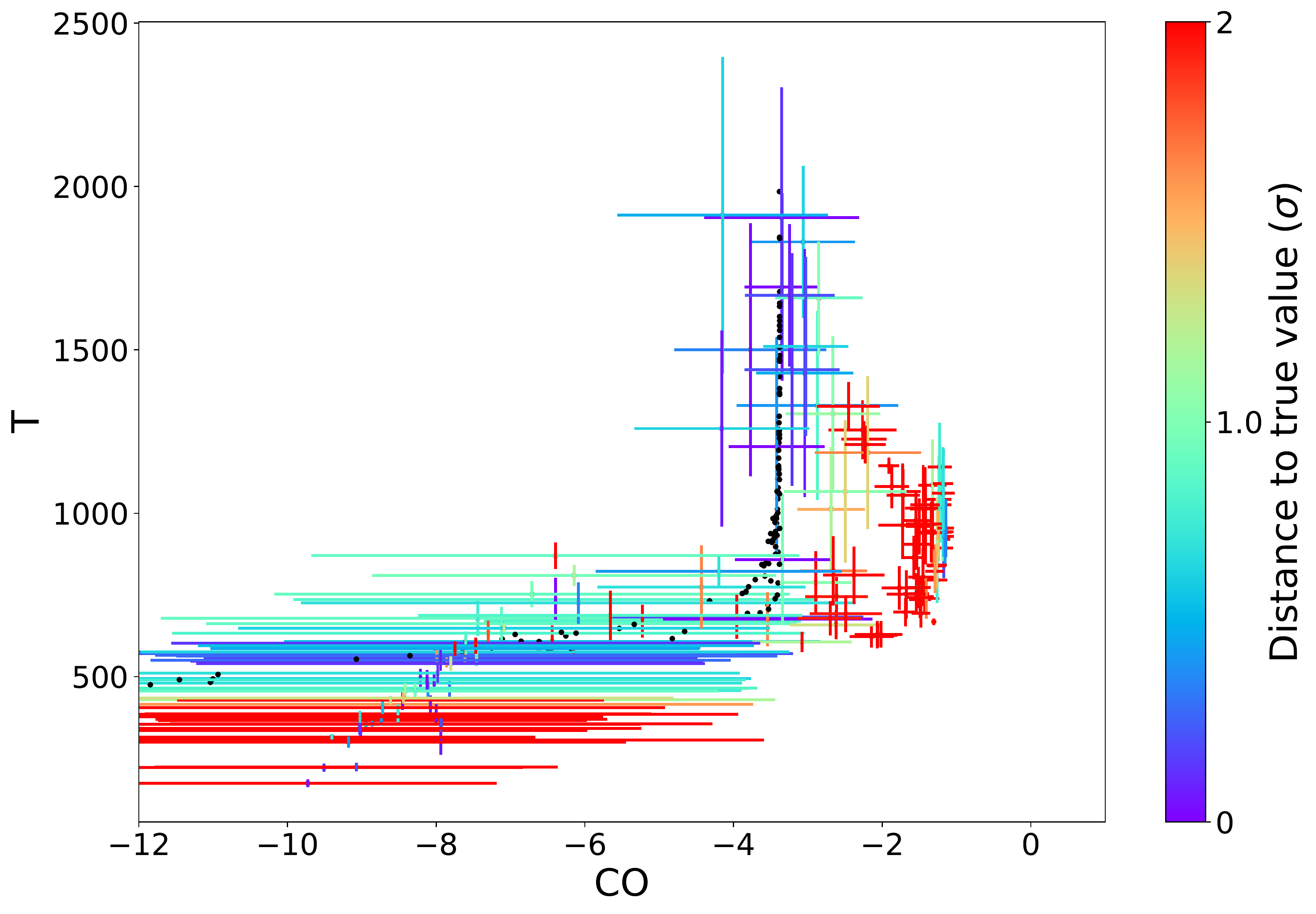}}
\caption{Biased sample:  equilibrium chemistry atmospheres.  Correlation map of the retrieved abundance of CO and the temperature, with the 1-$\sigma$ retrieved error bars. Results obtained with equilibrium chemistry retrievals (top) and with free, constant with pressure chemistry retrievals (bottom). The colour-scale represents the distance to the true value in units of  1-$\sigma$.
}
\label{fig:aceCO}
\end{figure*}

\begin{figure*}[h]
\centering
\centerline{\includegraphics[width=0.81\textwidth]{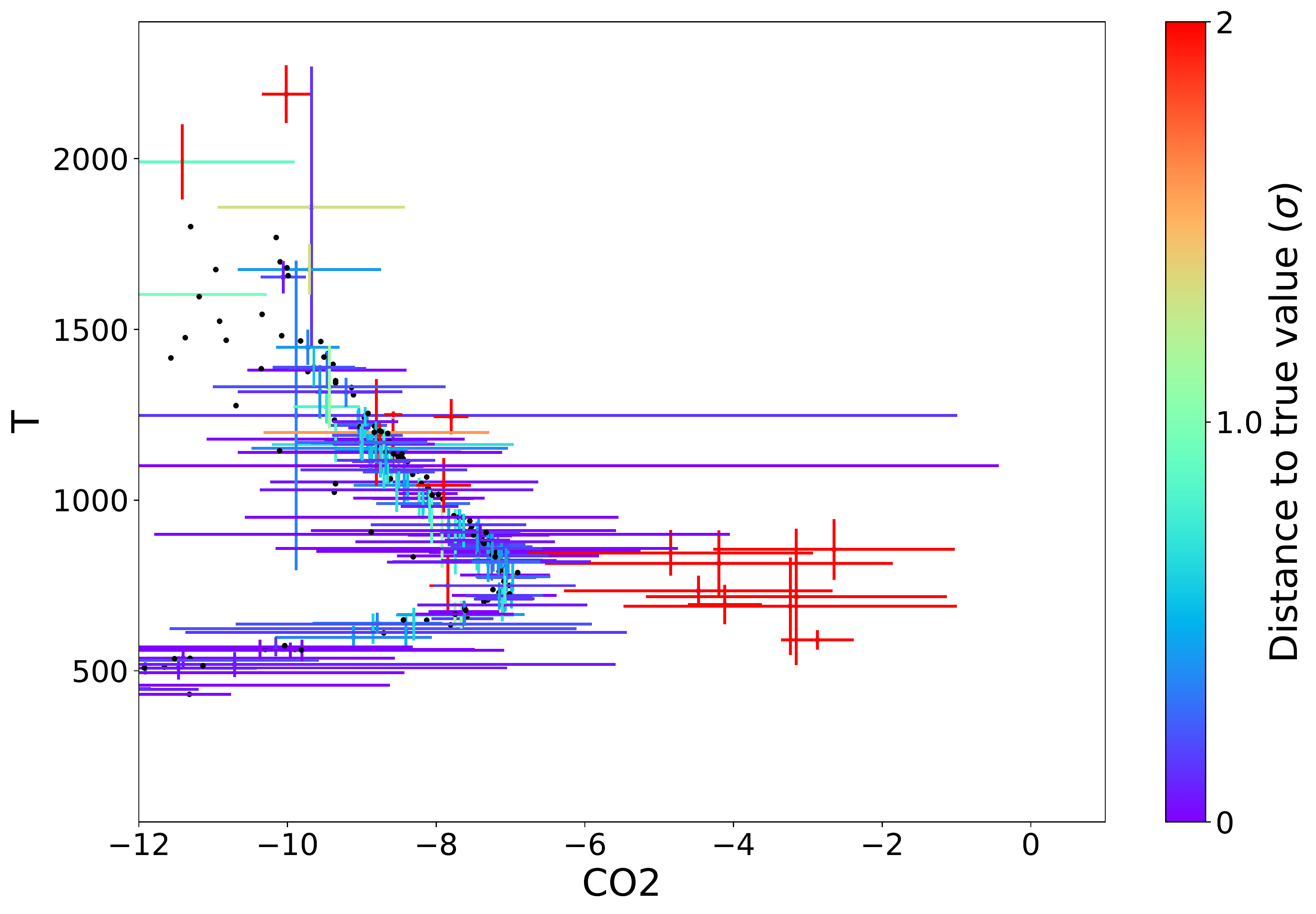}}
\centerline{\includegraphics[width=0.81\textwidth]{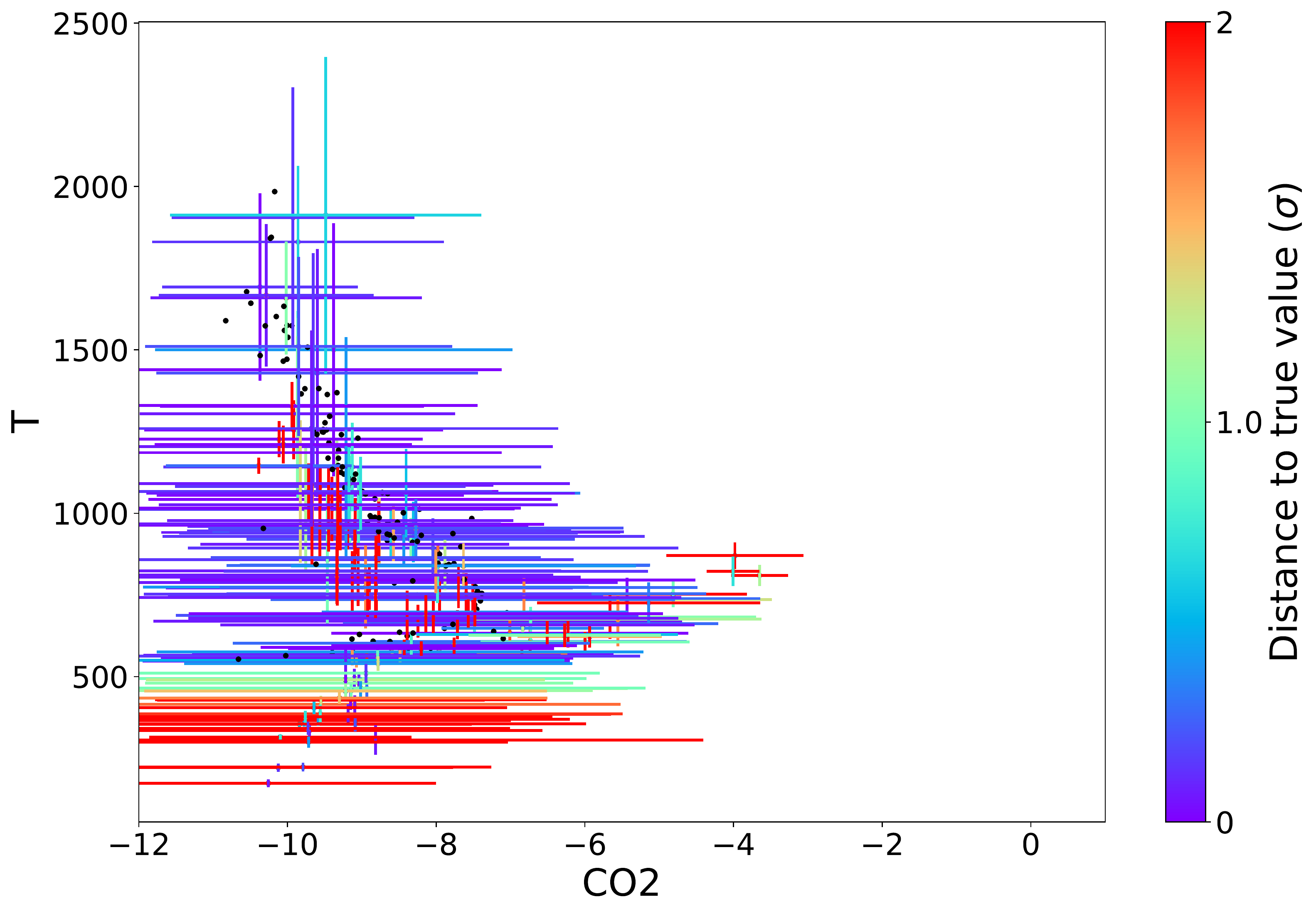}}
\caption{Biased sample:  equilibrium chemistry atmospheres.  Correlation map of the retrieved abundance of CO$_2$ and the temperature, with the 1-$\sigma$ retrieved error bars. Results obtained with equilibrium chemistry retrievals (top) and with free, constant with pressure chemistry retrievals (bottom). The colour-scale represents the distance to the true value in units of  1-$\sigma$.
}

\label{fig:aceCO2}
\end{figure*}

\end{document}